\documentclass[useAMS,usenatbib]{mn2ee}
\usepackage{amsmath,graphics,color,subfigure,epsfig,amssymb}
\usepackage{mathptmx}
\usepackage{epsfig}
\usepackage{booktabs}
\usepackage{rotating,rotfloat}
\usepackage{longtable}
\usepackage{graphicx}
\usepackage{url} 
\usepackage{listings}
\usepackage{color}
\usepackage{pslatex} 
\usepackage{booktabs}
\usepackage{dcolumn}

\usepackage[hyperindex,breaklinks=true, colorlinks, citecolor=blue]{hyperref} 

\def\ba{\begin{eqnarray}}
\def\ea{\end{eqnarray}}
\def\be{\begin{equation}}
\def\ee{\end{equation}}

\def\bdelta{{\delta }}

\newcommand{\transpose}[1]{\ensuremath{#1^{\scriptscriptstyle T}}}
\makeatletter
\def\ref@jnl#1{{\rmfamily#1}}%
\newcommand\aj{\ref@jnl{AJ}}%
\newcommand\araa{\ref@jnl{ARA\&A}}%
\newcommand\apj{\ref@jnl{ApJ}}%
\newcommand\apjl{\ref@jnl{ApJ}}%
\newcommand\apjs{\ref@jnl{ApJS}}%
\newcommand\apss{\ref@jnl{Ap\&SS}}%
\newcommand\aap{\ref@jnl{A\&A}}%
\newcommand\aapr{\ref@jnl{A\&A~Rev.}}%
\newcommand\aaps{\ref@jnl{A\&AS}}%
\newcommand\baas{\ref@jnl{BAAS}}%
\newcommand\memras{\ref@jnl{MmRAS}}%
\newcommand\mnras{\ref@jnl{MNRAS}}%
\newcommand\pra{\ref@jnl{Phys.~Rev.~A}}%
\newcommand\prb{\ref@jnl{Phys.~Rev.~B}}%
\newcommand\prc{\ref@jnl{Phys.~Rev.~C}}%
\newcommand\prd{\ref@jnl{Phys.~Rev.~D}}%
\newcommand\pre{\ref@jnl{Phys.~Rev.~E}}%
\newcommand\prl{\ref@jnl{Phys.~Rev.~Lett.}}%
\newcommand\pasp{\ref@jnl{PASP}}%
\newcommand\pasj{\ref@jnl{PASJ}}%
\newcommand\ssr{\ref@jnl{Space~Sci.~Rev.}}%
\newcommand\nat{\ref@jnl{Nature}}%
\newcommand\iaucirc{\ref@jnl{IAU~Circ.}}%
\newcommand\aplett{\ref@jnl{Astrophys.~Lett.}}%
\newcommand\apspr{\ref@jnl{Astrophys.~Space~Phys.~Res.}}%
\newcommand\nphysa{\ref@jnl{Nucl.~Phys.~A}}%
\newcommand\physrep{\ref@jnl{Phys.~Rep.}}%
\newcommand\planss{\ref@jnl{Planet.~Space~Sci.}}%
\newcommand\procspie{\ref@jnl{Proc.~SPIE}}%

\makeatletter

\makeatletter
\newcommand\footnoteref[1]{\protected@xdef\@thefnmark{\ref{#1}}\@footnotemark}
\makeatother

\title[Simulations for single-dish intensity mapping experiments]{Simulations for single-dish intensity mapping experiments}

\author[M.-A.\,Bigot-Sazy et al.]
  {M.-A.~Bigot-Sazy,$^1$\thanks{E-mail: marie-anne.bigot-sazy@manchester.ac.uk}
  C.~Dickinson,$^1$\thanks{E-mail: clive.dickinson@manchester.ac.uk} R.A.~Battye,$^1$ I.W.A.~Browne,$^1$
   \newauthor 
  Y.-Z.~Ma,$^1$ 
  B.~Maffei,$^1$
  F.~Noviello,$^1$
   M.~Remazeilles,$^1$
  P.N.~Wilkinson$^1$ \\
  $^1$Jodrell Bank Centre for Astrophysics, Alan Turing Building, School of Physics \& Astronomy, The University of Manchester, Oxford Road, \\
Manchester M13 9PL, UK \\
 }

\begin{document}


\setlength{\topmargin}{-15mm}

\pagerange{\pageref{firstpage}--\pageref{lastpage}} \pubyear{2002}

\maketitle

\label{firstpage}

\begin{abstract}
HI intensity mapping is an emerging tool to probe dark energy. Observations of the redshifted HI signal will be contaminated by instrumental
noise, atmospheric and Galactic foregrounds. The latter is expected to be four orders of magnitude brighter than the HI emission we
wish to detect. We present a simulation of single-dish observations
including an instrumental noise model with 1/$f$ and white noise, and
sky emission with  a diffuse Galactic foreground and HI emission. We
consider two foreground cleaning methods: spectral parametric fitting and principal component
analysis.  For a smooth frequency spectrum of the foreground and instrumental effects, we find that the parametric fitting method
provides residuals that are still contaminated by foreground and 1/$f$
noise, but the principal component analysis can remove this
contamination down to the thermal noise level. This method is robust for a range of different models of foreground and noise, and
so constitutes a promising way to recover the HI signal from the
data. However, it induces a leakage of the cosmological signal into
the subtracted foreground of around 5\%. The efficiency of the component
separation methods depends heavily on the smoothness of the frequency
spectrum of the foreground and the 1/$f$ noise. We find that as,
long as the spectral variations over the band are slow compared to the
channel width, the foreground cleaning method still works.
\end{abstract}

\section{Introduction}\label{sec:intro}
One of the main challenges of modern cosmology is to explain the
late-time acceleration of the expansion of the Universe. There are
several independent methods to probe dark energy: Baryonic Acoustic
Oscillations (BAOs), weak and strong gravitational lensing, cluster
counts and supernovae. However, BAO measurements appear to be the most
powerful cosmological tool at low redshift because they are limited by
statistical rather than systematics errors. Using the BAOs as a
standard ruler measures the expansion of the Universe as a function of
redshift, and so constrains the properties of dark energy (e.g. \citealp{Weinberg2013}).

A complementary method to the usual large optical surveys of galaxies
to study BAOs is HI intensity mapping \citep{Battye2004,Peterson2006,Chang2008,Loeb2008,Peterson2009}. This method
aims to give a tomographic distribution of the neutral HI emission present in the
recent Universe over large angular scales. Simulations indicate that
the HI intensity mapping technique will give very precise constraints
on cosmological parameters, and in particular, on the dark energy
equation of state at low redshift \citep{Chang2008}, and at high
redshift \citep{McQuinn2006,Bowman2007,Mao2008}. This sensitivity
comes from the large volume of the survey. 

Some HI experiments are
currently underway such as
BAOradio\footnote{http://arxiv.org/pdf/1209.3266v1.pdf},
BINGO\footnote{http://www.jb.man.ac.uk/research/BINGO/},
CHIME\footnote{http://chime.phas.ubc.ca/},
FAST\footnote{http://fast.bao.ac.cn/en/},
TIANLAI\footnote{http://tianlai.bao.ac.cn/}. Using intensity mapping, the Green Bank
Telescope (GBT)\footnote{https://science.nrao.edu/facilities/gbt/} has
provided the first detection of the HI signal at $z \sim 0.8$
cross-correlated with the WiggleZ Dark Energy Survey
\citep{Masui2013}. This detection shows the HI intensity mapping as a
promising tool and gives a lower limit on the fluctuations power of
the HI signal.  The first phase of the SKA instrument will be built in the
next decade and it will offer a broad range of frequencies and a
large survey area. Thus, this instrument has great potential to
deliver maps of the HI intensity \citep{Bull2015}.

To convert the promise into reality radio observations will have to
deal with different contaminants, which can dominate the signal of
interest in the data, such as astrophysical foregrounds, radio
frequency interference (RFI) and instrumental
noise. Another important contaminant is time variable noise introduced
during propagation of the signal through the atmosphere, which gives an additional contribution to the 1/$f$ noise of the instrument. The
amplitude of the atmospheric effects depends on the observing
frequency, on the elevation of the instrument above sea level and on the instantaneous weather conditions.

The most important challenge for any intensity mapping experiment is
the control of foreground emissions. Thus, the data analysis must
include a robust foreground subtraction algorithm. At $\sim$ 1 GHz,
the most relevant foregrounds are a combination of Galactic emission,
mostly synchrotron, and that from the background of extragalactic
point sources. See, for example, the discussion in \citet{Battye2013}. These emissions are at least four orders of magnitude
larger ($T_\textrm{b} \sim 1000$\,mK) than the HI signal fluctuations
($\delta T_\textrm{b} \sim 1$\,mK). In order to subtract the
foreground, the high spectral resolution offered by any HI experiment
allows us to exploit the frequency information. In particular the
spectra of the foregrounds are expected to be smooth and can be
approximated over the frequency range of interest to first order by a modified
power-law with a spatial curvature as a function of frequency \citep{Kogut2012}. This spectral smoothness
can be used to separate the HI signal from foreground signals. The
most common approach is to fit a smooth function to the data in the
frequency space and remove it. 

Several component separation techniques
have been discussed in the literature for removal of the Galactic
foreground, both specific to low redshift intensity mapping and to
epoch of reionisation experiments. We can classify the foreground
cleaning methods into two different categories: parametric and blind methods. The parametric methods are model-dependent and consist of 
applying a parametric fitting to each pixel in the maps \citep{Ansari2012}. The blind methods do
not require any assumption on the properties of the foregrounds and
the instrument response. Examples of such methods are FASTICA
\citep{Chapman2012,Wolz2014}, Correlated Component Analysis (CCA) method \citep{Bonaldi2014},  Karhunen-Loeve Decomposition \citep{Shaw2014}, GMCA \citep{Chapman2012}, principal
component analysis (PCA) and independent component analysis (ICA)
\citep{Alonso2015}. All methods must also deal with instrumentally induce effects such as mode-mixing of angular and frequency
fluctuations induced by the dependence of the beam with
frequency. The component separation methods are based on the spectral smoothness of the foregrounds, so the calibration will be a critical step in order not to compromise this smoothness. 

There are two instrumental approaches to HI mapping, either using
single dishes with multiple feeds or interferometer arrays. The
single-dish approach offers a relatively cheap and simple way for
doing intensity mapping. Unlike single-dish experiments which
naturally have good surface brightness sensitivity, interferometer
arrays suitable for intensity mapping will require many close-packed
elements in order to detect the very low surface brightness HI signal
and thus need big correlators \citep{thompson2008interferometry}. Both kinds of experiment will have to
deal with potential systematics similar to those encountered in cosmic microwave background (CMB) imaging experiments, such as gain
variations, spillover, ground pickup and correlated noise in
frequency. Single-dish experiments will require stable receiver
systems, good calibration and an appropriate scanning strategy. On the
other hand, interferometers are known to deal more naturally with
systematics and with foregrounds than single dishes and hence
receiver stability (e.g. \citealp{Dickinson2004,Readhead2004}), etc. is not such an important issue. However, we
point out that existing interferometers are limited by the small
number of their smallest baselines and hence fail to provide the
required surface brightness sensitivity \citep{Bull2015}.

In this paper, we focus on the concept of a single-dish experiment
with the BINGO (BAO from Integrated Neutral gas Observations)
instrument, which aims at mapping the HI emission at frequencies from
960 to 1260\,MHz ($z= 0.12-0.48$). This experiment will measure the
HI power spectrum, and will detect for the first time the Baryon
Acoustic Oscillations around 1 GHz. Some of the details of the BINGO
experiment can be found in \citet{Battye2013}. Though we use the BINGO
instrument as a concrete example, nearly all the following analysis
concepts can be applied to other single-dish instruments.

We organise the paper as follows. In Section~\ref{sec:simu}, we
describe our simulations in which we incorporate foreground and
instrumental noise models. We also make some predictions of the total
atmospheric contribution to the instrumental noise level and of the
atmospheric fluctuations coming from the inhomogeneous distribution of
water vapour. In Section~\ref{sec:fg_noise_sep}, we focus on two
simple foreground and noise subtraction procedures: parametric fitting
and principal component analysis. The
success of these methods depends on the smoothness of the frequency
spectrum of the noise and the foreground. In
Section~\ref{sec:smoothness}, we place some requirements on the
smoothness of these spectra needed for extracting the HI signal from
the instrumental $1/f$ noise and from the brighter foreground, the
synchrotron emission. In this way, we assess the robustness of the
cleaning methods, according to the smoothness of the frequency
spectrum of the simulated data.
\section{Simulation of a single-dish experiment}\label{sec:simu}

In order to explore the foreground cleaning methods we use simulations of the proposed
BINGO telescope \citep{Battye2013} as a concrete example of a single-dish instrument. To do this we require simulated maps of the sky at
the observation frequencies. We produce a time-ordered data stream
with a foreground model which includes Galactic synchrotron plus a
background of unresolved point sources, detailed in
Section~\ref{subsec:fg_em}, while the way we produce the HI signal is
described in Section~\ref{subsec:21cm}. In
Section~\ref{subsec:mapmaking}, we introduce the map-making method
used to obtain the sky maps of the experiment and describe the instrumental
noise model ($1/f$ and thermal noise) in
Section~\ref{subsec:instru_noise}. Finally, in
Section~\ref{subsec:atm_prediction}, we make some predictions of the
amplitude of the atmospheric noise.

\subsection{Foreground model}\label{subsec:fg_em}

\subsubsection{Galactic synchrotron}\label{subsubsec:synch}
To generate a template of the sky emission, we use the reprocessed 408\,MHz Haslam et al. map \citep{Remazeilles2014}, which constitutes a
good tracer of the diffuse synchrotron emission. The synchrotron
spectrum in terms of brightness temperature can be approximated by $T(\nu) \propto
\nu^{\beta+C\text{\textnormal{ln}}(\nu/\nu_0)}$ \citep{Kogut2012}, where $\nu$ is the
radiation frequency, $C$ the curvature defined with respect to a
reference frequency $\nu_0$ and $\beta$ is the spectral index at $\nu=\nu_0$. Observations have indicated that there are spatial
variations of the spectral index \citep{Reich1988,Platania1998,Davies2006}. We extrapolate this template to the frequencies of
interest by using 3 different models for $\beta$ listed
below, from the simplest to the more complicated ones: \\ 1. We ignore any
variation across the sky of the spectral index $\beta$, we fix this index to an average value estimated at frequencies near $\sim 1$ GHz
$\beta=-2.8$ \citep{Platania1998}. \\ 2. We assume a Gaussian spatial
distribution of the synchrotron index $\beta$, with $<\beta>=-2.8$ and
a r.m.s. value of 0.17 \citep{Platania1998}. \\ 3. We use spectral
model of the global sky from 10\,MHz to 100\,GHz developed by
\citet{deOliveiraCosta2008}. This final model is the most realistic one. It includes spatial correlations of the Galactic emission across the sky
and a frequency curvature modification of the synchrotron index $\beta$. The mean value of $\beta$ is $-2.5$ and the steepening of this index is 0.03.

The models listed above are summarised in Table \ref{tab:synchrmodel}.

\begin{table}
\caption{Summary of the different models of the Galactic synchrotron emission.}
  \label{tab:synchrmodel}
  \begin{center}
    \leavevmode
    \setlength{\tabcolsep}{3pt}
    \begin{tabular}{lclclr} \hline \hline              
   \small Synchrotron  &  Characteristics 										   & Mean $\beta$   & r.m.s. $\beta$               \\ \hline 
   \small Model 1	              & \small $\beta$ constant on the sky 					   & $-2.8$     		       & $\beta$ constant    	 \\   
   \small Model 2                      & \small A Gaussian spatial distribution of $\beta$  	      	   & $-2.8$    		       &	0.17				 \\ 
   \small Model 3                      & \small \citet{deOliveiraCosta2008} model                   		   & $-2.5$                &	       0.03				\\
  \hline
       \end{tabular}
  \end{center}
\end{table}

A high-resolution template of synchrotron emission is required to make
realistic tests of foreground removal methods. However, the resulting
synchrotron map has a resolution corresponding to a beam with FHWM
equal to 56\,arcmin. We require a higher resolution map so it cannot be
directly used as a template of the synchrotron emission. And, there are
no other full-sky astronomical data sets with
resolution better than $\sim$ 1$^{\circ}$. Hence, to account for these small-scale
fluctuations, we add to the original map a random Gaussian realisation
with a power spectrum $C_{\ell}=
\ell^\gamma(\text{\textnormal{exp}}(-\ell^2\sigma^2_{{\textrm{sim}}}))$,
where $\gamma=-2.7$, $\sigma_{{\textrm{sim}}}$ is the Gaussian width of
the simulation and $\ell$ the multipole. The details are given in
\citealt{MivilleDeschenes2007} and \citealt{Remazeilles2014}.

\subsubsection{Extragalactic point sources}\label{subsubsec:ps}
We assume that the distribution of such sources is not spatially
correlated, that is to say the clustering is weak
\citep{Liu2009} and hence that they are Poisson distributed. The clustering part increases the pixel-pixel correlations \citep{Battye2013} and thus can have an impact on the foreground removal method. In subsequent work, we will investigate this contribution. Extragalactic point sources can be divided into two
populations. The first component comprises bright and isolated point
sources that can be readily detected by the instrument and removed directly using the data of the experiment. The second population consists of a continuum of unresolved sources.

At radio frequencies (GHz), the r.m.s. confusion $\sigma_{\textrm{c}}$ in a telescope beam with the full width at half maximum $\theta_{\textrm{FWHM}}$ can be approximated by \citep{Condon1974}
\begin{equation}
\frac{\sigma_{\textrm{c}}}{\text{\textnormal{mJy}}} \approx 0.2
\left( \frac{\nu}{\text{\textnormal{GHz}}} \right)^{-0.7} \left(
\frac{\theta_{\textrm{FWHM}}}{\text{\textnormal{arcmin}}} \right)^{2}.
\end{equation}
For the BINGO telescope, with a $\theta_{\textrm{FWHM}}=40$\,arcmin, this is around 320\,mJy at 1000\,MHz, thus BINGO will be subject to confusion noise when considering a continuum detection. This is irrelevant for an HI line signal. 

The brightness of each source is drawn from the differential source counts $\frac{\textrm{d}N}{\textrm{d}S}$, with the number of sources per steradian $N$ and per unit of flux $S$. In \citet{Battye2013}, they use data from multiple continuum surveys at 1.4\,GHz \citep{Mitchell1985, White1997,Ciliegi1999, Gruppioni1999, Hopkins1999, Richards2000, Bondi2003, Fomalont2006,  Owen2008, Seymour2008,Ibar2010} and fit a 5th order polynomial to these data

\begin{equation}
\textnormal{log}_{10}\left( \frac{S^{2.5}\textrm{d}N/\textrm{d}S}{N_0}\right)=\sum_{i=0}^5a_i\begin{bmatrix}\textnormal{log}_{10}\left( \frac{S}{S_0}\right)\end{bmatrix}^i,
\end{equation}
where $a_0=2.593$, $a_1=9.333\times 10^{-2}$, $a_2=-4.839\times10^{-4}$, $a_3=2.488\times10^{-1}$, $a_4= 8.995\times10^{-2}$ and $a_5=8.506\times10^{-3}$; and $N_0 = 1$\,Jy$^{3/2}$\,sr$^{-1}$ and $S_0 = 1$\,Jy. The power-law spectral function with a Gaussian distributed index is given by 
\begin{equation}
S(\nu)=S(1.4 \, \text{\textnormal{GHz}})\left( \frac{\nu}{1.4 \, \text{\textnormal{GHz}}}\right)^{-\alpha}.
\end{equation}
The spectral index $\alpha$ is randomly chosen from a Gaussian distribution 
\begin{equation}
P(\alpha)=\frac{1}{\sqrt{(2\pi \sigma^2)}}\text{\textnormal{exp}}\begin{bmatrix}-\frac{(\alpha-\alpha_0)^2}{2\sigma^2}\end{bmatrix},
\end{equation}
with a mean of  $\alpha_0=2.5$ and a width distribution of $\sigma=0.5$ \citep{Tegmark2000}.

Assuming that the sources with flux $S > S_{\textnormal{max}}$ can be subtracted from the data, we estimate the mean brightness temperature, contributed due to the remaining sources, by  
\begin{equation}
T_{{\textrm{ps}}}(\nu,\hat{n})=\left( \frac{\textrm{d}B}{\textrm{d}T} \right)^{-1}\Omega_{\textrm{pix}}^{-1}\sum_{i=1}^{N}S_i(\nu),
\end{equation} 
where $S_i$ is the flux of the point source $i$ at 1.4 GHz and $ \Omega_{\textnormal{sky}}$ the pixel size qual to 0.22 arcmin$^2$. The parameter
$\textrm{d}B/\textrm{d}T=2k_\textrm{B}/\lambda^2$ is the conversion
factor between intensity units to brightness temperature units,
$k_{\textrm{B}}$ being the Boltzmann constant, $\lambda$ the
wavelength of the incoming radiation and $\Omega_{\textrm{pix}}$ the pixel size.

We expect that the brightest sources could be removed directly from the BINGO data or could be masked using the NRAO VLA Sky-Survey (NVSS) \citep{Condon1998}, which is considered to be 99\% complete at a flux density limit of 3.4\,mJy. For our simulation, to be conservative, we take $S_{\textnormal{max}}=100$\,mJy, which corresponds to $\sim 1$ source per square degree. We expect to either subtract or mask most of the brightest radio sources above this flux density. We will investigate in a following paper the residual contribution due to the variability of radio sources and calibration issues.

\begin{figure}
  \begin{center}
   \includegraphics[width=\columnwidth]{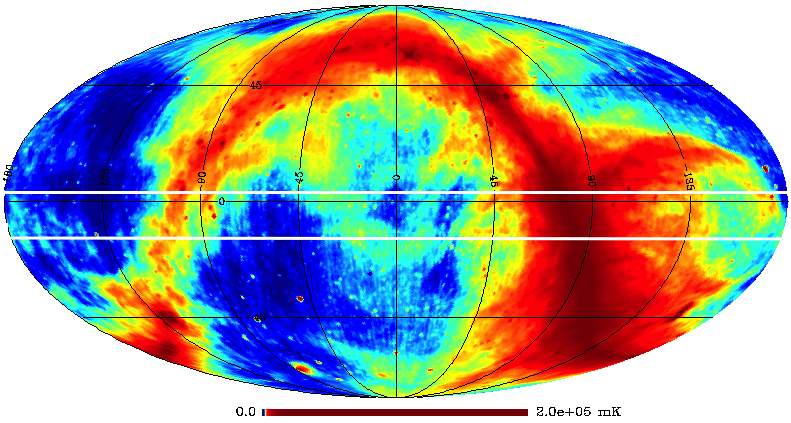}
\end{center}
\caption{Mollweide projection of the foreground with a background of unresolved point sources (S\,$<100$\,mJy) and synchrotron emission at 1000\,MHz in celestial (RA/Dec) coordinates with RA=0$^\circ$ at the centre and increasing to the left. The white solid lines define the spaces expected to be observed by the BINGO experiment. }
\label{Fig:fgmap}
\end{figure}

In the following, the maps are created using the HEALPix pixelisation scheme \citep{Gorski2005}. A foreground map of this simulation at 1000\,MHz is given in
Fig.~\ref{Fig:fgmap}. The colour bar represents the brightness
temperature in mK.

\subsection{HI signal}\label{subsec:21cm}

We use the CORA software \citep{Shaw2014} to simulate the HI
brightness temperature. We assume the Planck+WP+highL+BAO cosmological model given in \citet{PlanckCollaboration2013}. The HI
brightness temperature can be written as a sum of two parts

\begin{equation}
T_{\textrm{b}}=\bar{T}_{\textrm{b}}(1+\delta_{{{\textrm{HI}}}}),
\end{equation}
where $\delta_{{{\textrm{HI}}}}$ is the HI density contrast and $\bar{T}_\textrm{b}$ the mean HI brightness temperature given by 
\small
\begin{align}
\bar{T}_\textrm{b}(z)= 0.3\textrm{K}\left( \frac{\Omega_{{\textrm{HI}}}}{10^{-3}}  \right) \left( \frac{\Omega_{\textrm{m}}+(1+z)^{-3}\Omega_{\Lambda}}{0.29} \right)^{-1/2} \left( \frac{1+z}{2.5} \right) ^{1/2}.
\end{align}
\normalsize
We assume that the neutral HI fraction is $\Omega_{{\textrm{HI}}}=5\times 10^{-4}$ \citep{Switzer2013} and the HI bias is independent of scale and redshift with $b_{{\textrm{HI}}}=1$. 

The HI brightness temperature power spectrum can be modeled as 
\begin{equation}
P_{{\textrm{T}}_{\textrm{b}}}(\vec{k},z)=\bar{T}^2_{\textrm{b}}(z) \begin{bmatrix} \textrm{b}_{\textrm{HI}}+f\mu^2\end{bmatrix}^2 D^2(z)P_{\textrm{m}}(k,z),
\end{equation}
where $\mu \sim k_{\parallel}/k$ with the flat-sky approximation, $P_{\textrm{m}}(k,z)$ the matter power spectrum, $D(z)$ the linear growth factor normalised by $D(0)=1$, and $f$ the linear growth rate $f=\text{\textnormal{d\,log}}D/\text{\textnormal{d\,log}}a,$ where $a$ is the cosmological scale factor. The HI angular power spectrum is obtained from Gaussian random fields with the flat sky angular power spectrum \citep{Datta2007}
\begin{equation}
C^{\textrm{flat}}_{\ell}(\Delta \nu)=\frac{\bar{T}_{\textrm{b}}^2}{\pi r^2_{\textrm{v}} }\int_0^\infty \textrm{d} k_\parallel  \text{\textnormal{cos}}(k_\parallel r_{\textrm{v}}  \Delta \nu)P_{{\textrm{T}}_{\textrm{b}}}(\textbf{k}),
\end{equation}
where $r_{\textrm{v}}$ is the comoving distance, $\textbf{k}$ has components $k_\parallel$ and $\ell/r_{\textrm{v}}$ along the line-of-sight and in the plane of the sky respectively.
Using these inputs, we generate the maps of the HI signal which have r.m.s. fluctuations around $ 0.1$\,mK.

\subsection{Simulation of a single-dish experiment}\label{subsec:mapmaking}
We consider a single-dish experiment based on the BINGO concept. BINGO will be
a dual mirror Compact Antenna Test Range (CATR) telescope with a 40\,m primary
mirror and an offset focus. Apart from the telescope optics the design of the instrument is
similar to that of \citet{Battye2013}. The proposed BINGO experiment will have a receiver array containing between 50 and 60 feed horns. In our simulation, we model the receiver plane
with 56 feed horns with a 90\,m focal length. We consider the frequency range from 960\,MHz ($z=0.48$)
to 1260\,MHz ($z=0.13$). To decrease the computational speed, we choose to divide the
300\,MHz band into 20 channels, each of 15\,MHz bandwidth, though the actual instrument will have much narrower frequency channels
to facilitate RFI excision. The sampling rate is 0.1\,Hz. The instrumental parameters used for
our simulation are listed in Table \ref{tab:survp}.

\begin{table}
\caption{Instrumental parameters for BINGO simulation.}
  \label{tab:survp}
  \begin{center}
    \leavevmode
    \begin{tabular}{lcr} \hline \hline              
   \small Survey parameters  &                     		\\ \hline 
   \small Redshift range [$z_{\textrm{min}}, z_{\textrm{max}}$]  			    & \small [0.13, 0.48]  	  \\    
   \small Frequency range [$\nu_{\textrm{min}}, \nu_{\textrm{max}}$]  (MHz) & \small [960, 1260]  	  \\ 
   \small Channel width $\Delta \nu$ (MHz)            		                                & \small 15                         \\
   \small FWHM (arcmin) at 1 GHz            			   	   			    & \small 40                         \\
   \small Number of feed horns $n_\textrm{f}$		                    		   	    & \small  56		           \\
   \small Sky coverage $\Omega_{{\textrm{sur}}}$ (deg$^2$)                	    & \small 3000                    \\
   \small Observation time $t_{{\textrm{obs}}}$ (yr)            		            	    & \small 1                         \\
   \small System temperature $T_{{\textrm{sys}}}$ (K)            		             & \small 50                         \\
   \small Sampling rate (Hz)	        			       		            		    & \small 0.1                   \\

   \hline
       \end{tabular}
  \end{center}
\end{table}

We assume that the horns are arranged in a rectangular configuration spaced 3.3\,m apart and the beams are given by a circular Gaussian. The beams are diffraction-limited and, therefore, the full width at half maximum $\theta_{\textrm{FWHM}}$ of the beam can be scaled to any frequency $\nu$ by
\begin{equation}
\theta_{\textrm{FWHM}}(\nu)=\theta_{\textrm{FWHM}}(\nu_0)\frac{\nu_0}{\nu},
\end{equation}
with $\nu_0=1000$\,MHz and $\theta_{\textrm{FWHM}}(\nu_0)=40$\,arcmin.

For the following simulations, we will assume that the telescope will map a $15\degr$ declination strip centred at $-5\degr$ as the sky drifts past the telescope. The declination of $-5\degr$ has been chosen to minimise the foreground emission, which is lowest between 10 and $-10\degr$ declination. We assume one full year of on-source integration. In practice, this will likely represent about 2 years of real observation time since we could consider only night observations we will probably remove some data due to foreseeable technical issues like such as radio frequency interference, weather downtime etc. 

In order to obtain the simulated BINGO maps, we use a maximum
likelihood map-making algorithm \citep{Stompor2002,Hamilton2003}. We
model the timelines $\bf{d}$ as $\textbf{d}=A\bf{s}+\bf{n}$, where $\bf{s}$ is the pixelized sky
signal which is mapped into the timelines and corrupted by noise
$\bf{n}$. The pointing information is represented by the pointing matrix
$A$ of size N$_{{\textrm{samples}}}$ $\times$ N$_{{\textrm{pixels}}}$,
which connects the time index to the pixel index. The map-making step
is given by
\begin{equation}
\hat{\textbf{s}}=({\transpose{A}} N^{-1}A)^{-1}{\transpose{A}}N^{-1}\textbf{d},
\end{equation}
 where $N$ is the noise covariance matrix and $\hat{\textbf{s}}$ is the best estimate of $\bf{s}$. An impact of the $1/f$ noise is to induce slow drifts of the gains of the receivers. If we do not take steps to
mitigate it, the $1/f$ noise will introduce stripes in the maps along
the direction of the drift scan. The inversion of $({\transpose{A}}
N^{-1}A)$ is performed by using the preconditioned conjugate gradient
method. The preconditioner is a pixel domain diagonal matrix weighting
the pixels by the number of times they have been observed. This method
is described in detail in \citet{Cantalupo2010}.

\begin{figure}
\centering
{
\includegraphics[height=0.4cm, width=\columnwidth]{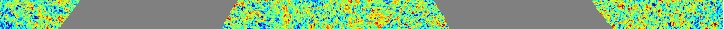}
}
\quad 
{
\includegraphics[width=5.5cm]{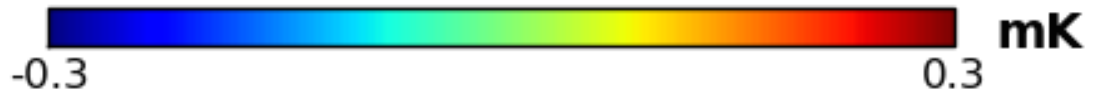}

}
\quad
{
\includegraphics[height=0.4cm, width=\columnwidth]{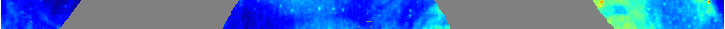}
}
\quad
{
\includegraphics[width=5.5cm]{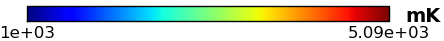}
}
\caption{Two drift scan strips of the observed sky with only HI signal
in the \textit{top} panel while the bottom panel includes Galactic synchrotron emission plus a background of unresolved point
sources. The mask of the Galactic plane at $|b|<20^{\circ}$ is plotted
with grey pixels. Note the different linear intensity scales. Colour bars represent the temperature in mK.}
 \label{Fig:rotsignal}
\end{figure}

We set the \texttt{HEALPix} resolution of the map equal to nside\,$=\,128$, which corresponds to a map pixel size of 27 arcmin. The focal plane configuration will lead to some gaps in the observed sky band. To correct for this, we rotate the beams of the horns on the sky with an angle $\sim5\,\degr$. In Fig.~\ref{Fig:rotsignal}, we show the drift scan strips of the sky
emission. In the following, we consider a single frequency channel
centered at 997.5\,MHz to display the results. The top panel shows the
HI signal and the bottom panel the Galactic synchrotron emission plus
a background of unresolved point sources. The amplitude of the
foreground emission is much higher than the signal of interest by
four orders of magnitude (note the difference in colour-scale in the
strips). These maps are plotted with a Cartesian projection using the
\texttt{HEALPix} software.

\subsection{Instrumental noise}\label{subsec:instru_noise}
The timelines are corrupted by thermal (white) noise. The optimal sensitivity of the BINGO experiment per pixel can be defined as follows
\begin{equation}
\sigma_{\textrm{t}}=\frac{T_{{\textrm{sys}}}}{\sqrt{ t_{{\textrm{pix}}}\Delta \nu }},
\label{eq:therm}
\end{equation}
where $\Delta\nu$ is the frequency channel given in Table \ref{tab:survp}. We assume the same system temperature $T_{{\textrm{sys}}}$ for all receivers. The parameter $t_{{\textrm{pix}}}$ is the integration time per pixel defined by
\begin{equation}
t_{{\textrm{pix}}}=n_{\textrm{f}}t_{{\textrm{obs}}}\frac{\Omega_{{\textrm{pix}}}}{\Omega_{{\textrm{sur}}}}, 
\end{equation}
where $n_{\textrm{f}}$ denotes the number of feed horns,
$t_{{\textrm{obs}}}$ is the total integration time, $\Omega_{{\textrm{sur}}}$ is the survey area and $\Omega_{{\textrm{pix}}}$ is the beam area. The
values of these parameters are given in Table \ref{tab:survp}.  We
assume $\Omega_{{\textrm{pix}}}=\theta_{\textrm{FWHM}}^2$ and for an
integration time of one year, we obtain $\sigma_{\textrm{t}}= \, 25\,
\mu$K.

Our simulation also contains $1/f$ noise, produced by gain fluctuations of the amplifiers
and thus this noise is correlated across all frequency channels. The impact of
these fluctuations is usually simulated in the frequency domain using
a 1/$f$ power spectrum. To generate realistic 1/$f$ noise we use
an algorithm to create a time sequence of white noise. Then we compute
its Fourier Transform and weight the data by a zero-mean power
spectral density with the distribution
\begin{equation}
P_{{\textrm{sd}}}=\frac{\sigma_{\textrm{t}}^2}{\nu_{{\textrm{samp}}}}\left[1+ \left( \frac{f_{{\textrm{knee}}}}{f} \right) ^{\alpha}\right],
\end{equation}
where $\nu_{{\textrm{samp}}}$ is the sampling frequency. $f$ is the discrete Fourier transform sample frequency,
given a length of the number of time samples and a sample spacing of
1/$\nu_{\textrm{samp}}$. The 1/$f$ knee
value is the integration time (the inverse of the discrete Fourier
transform sample frequency) at which the thermal and $1/f$ noise make
equal contributions to the power spectral density. Finally, we compute the inverse Fourier
transform of these data to obtain the time-ordered data of the noise.

In practice, we will filter the data on timescales of 20\,min during data processing; this is the timescale for which the largest structures of interest take to drift through the BINGO field-of-view. For simplicity, we simulate each 20\,min timestream separately and join them together by fixing the first sample of the nth equal to the $(n-1)$ sample. We assume the same
value of the knee frequency for each receiver $f_{{\textrm{knee}}}=10^{-3}$\,Hz for a 15 MHz channel bandwidth, which corresponds to the value we aim to achieve with
the BINGO pseudo-correlation receivers. The 1/$f$ slope index $\alpha$ is
assumed to be 1. In this paper, we start by making the assumption
that the $1/f$ noise is perfectly correlated between the frequency
channels, which is what is expected if we are dealing with simple gain
fluctuations and we assume a flat frequency spectrum ($\beta=0$). In Section~\ref{sec:smoothness}, we investigate what happens if these assumptions are relaxed.

\begin{figure*}
\centering
{
\includegraphics[width=15cm]{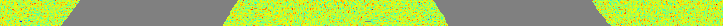}
}
\quad
{
\includegraphics[width=5.5cm]{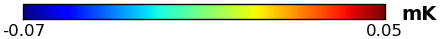}
}
\quad
{
\includegraphics[width=15cm]{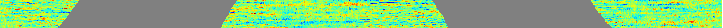}
}
\quad
{
\includegraphics[width=5.5cm]{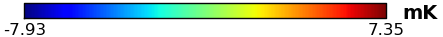}
}
\caption{Maps of the drift scan of the instrumental noise. The \textit{top} panel represents the thermal noise and the \textit{bottom} panel the $1/f$ noise. Note the different intensity scales. Colour bars represent the temperature in mK.} 
  \label{Fig:noisemap}
\end{figure*}

In Fig.~\ref{Fig:noisemap}, we show two maps of the noise with
thermal noise only (in the top panel) and with added $1/f$ noise (in
the bottom panel). One can notice the  stripes along the
direction of the scan induced by the instrumental $1/f$ noise, which is much larger than the thermal noise by a factor of
$\sim$ 100. 

\subsection{Atmospheric noise}\label{subsec:atm_prediction}

The observations of a ground-based telescope are affected by the
atmosphere at different levels, depending on the observing frequency
and of course on the prevailing weather conditions. The incoming
signal will be absorbed or scattered by the atmosphere. This effect
increases the noise level of the instrument and we quantify this in
Appendix~\ref{subsec:totalatm}. Around 1\,GHz, the optical depth is
dominated by oxygen which is constant in time with a small
contribution from water, usually quantified in terms of the precipitable
water vapour (PWV). Hence, it can undergo changes on an hourly (or faster) timescales and vary spatially. In Appendix~\ref{subsec:totalatm} we show
that the amount of PWV at 1\,GHz does not have a significant impact on
the total brightness temperature of the atmosphere. We find an
atmospheric contribution $\sim 1.81$\,K in days with favourable
atmospheric conditions PWV $< 2$\,mm, and $\sim 1.82$\,K in days with unfavourable weather PWV $> 4$\,mm. Thus, we expect the atmospheric contribution to the system
temperature to be at the level of a few percent.

However, we are also concerned with the fluctuating part of the emission and
absorption, which leads to an additional 1/$f$ noise-like component in
the time-ordered data. We quantify this effect in
Appendix~\ref{subsec:varatm}. The main source of fluctuations is
atmospheric turbulence in the troposphere. Oxygen molecules are
uniformly mixed in the atmosphere, but the water vapour molecules show
inhomogeneities. Even if the water contributes a small
percentage to the noise level, around 1\%, its variations with time
can produce fluctuations that are large compared to the HI
signal and thermal noise. This can generate some correlations between the time streams of data from different receivers and for the same receiver at different
times. These fluctuations depend on the weather conditions at the observing site, on the telescope, on the frequency bands and on the
observing strategy \citep{Church1995,Lay2000}. The effects of atmospheric noise have been well studied in cosmic microwave background experiments (e.g.
\citealp{Davies1996, Sayers2010}), and can be approximated by $1/f$ noise
at low-frequency. However, the amplitude of the atmospheric
fluctuations in the time streams has not been measured at 1\,GHz. From
the back-of-the-envelope calculation detailed in
Appendix~\ref{subsec:varatm}, we find that the amplitude of the
atmospheric fluctuations in antenna temperature for a single dish is
$\Delta T_{{\textrm{atm}}} \sim 0.01$\,mK. Thus, the amplitude of these
fluctuations is below the instrumental noise, which is expected to be $\sim$1\,mK for a frequency resolution of 15\,MHz for BINGO. So, the contribution from the atmospheric fluctuations appears
not to be a challenge for a single-dish experiment observing around 1\,GHz when the weather is stable.

\section{Foreground and instrumental noise subtraction}\label{sec:fg_noise_sep}

For an observing frequency around 1\,GHz, the synchrotron emission and
the extragalactic point sources are the most relevant
foregrounds. The removal of the foregrounds and instrumental $1/f$
will rely on the smoothness of their frequency
spectra. In this section, we want to quantify how well the foregrounds
can be subtracted in the presence of thermal and 1/$f$ noise. Our
philosophy is to focus on two simple cleaning procedures, parametric fitting and a blind method with principal component analysis (PCA). We
describe these methods in Section~\ref{subsec:sep_meth} and present
their results in Section~\ref{subsec:res_sep_meth}. We also
demonstrate the possibility of using a blind method to remove the
instrumental $1/f$ noise in Section~\ref{subsec:noise_sep}. In the
following, we assume no systematics and a perfect calibration of the data.

\subsection{Methods}\label{subsec:sep_meth}

\subsubsection{Parametric fitting}
Parametric fitting is a common method to parameterise foregrounds
(e.g. \citealp{Brandt1994, Ansari2012}). The approach of the method is to fit directly an explicit parametric model of the foregrounds and noise to each pixel of the maps along the frequency direction. The common foreground model is a modified power-law. As the main
foreground emission, the Galactic synchrotron can be approximated by
a parametric distribution with a curvature to first order \citep{Kogut2012}.

The $i$-th pixel of the simulated map of the sky at the frequency
$\nu_j$ can be written as the sum of the intensity of the HI signal
$T^i_{\textrm{21cm}}$, the foreground emissions $T^i_{\textrm{fg}}$
and the noise of the instrument $T^i_{\textrm{n}}$
\begin{equation}
\hat{T}_j^{i}=\hat{T}^i_{\textrm{21cm},j}+\hat{T}^i_{\textrm{fg},j}+\hat{T}^i_{\textrm{n},j}.
\end{equation}
The hat symbol denotes a modelled quantity. We make the assumption
that the foreground $\hat{T}_{\textrm{fg}}$ and $1/f$ noise $\hat{T}^i_{\textrm{n}}$ can be modelled
by
\begin{equation}
\hat{T}_{\textrm{fg},j}^{i}+\hat{T}^i_{\textrm{n},j}+=A^i\left(\frac{\nu}{\nu_0}\right)^{\beta},
\label{eq:plfit}
\end{equation}
where $\beta$ is the spectral index and $A$ is the amplitude in mK. This assumption on the spectral slope of the $1/f$ noise can be justified by the fact that the $1/f$ noise fluctuations are expected to have a spectral form similar to the system temperature, which can be approximated by a power-law over the BINGO frequency range. We fit Eq.~\ref{eq:plfit} for each pixel of the map in the frequency
direction minimised using a least-squares method.

\subsubsection{Principal Component Analysis (PCA)}

PCA  \citep{Murtagh1987} has the
advantage of being a non-parametric method and so requires no
specific prior information on the spectra of the foreground and the
noise. This method consists of transforming the independent maps of each
frequency channel into orthogonal modes according to the covariance between
frequencies.
\noindent
We consider the data to be a matrix $S$, with $N_f \times
N_p$ elements. $N_f$ denotes the number of frequency
channels and $N_p$ the number of pixels in the map. We compute the
frequency covariance matrix from the simulated data

\begin{equation}
C_{ij}=\frac{1}{N_p}S\transpose{S}=\frac{1}{N_p}\sum_{p=1}^{N_p}T(\nu_i,\hat{n}_p)T(\nu_j, \hat{n}_p), 
\end{equation}
where $T(\nu_i,\hat{n}_p)$ is the brightness temperature along the
direction of the line-of-sight $\hat{n}_p$ and for the frequency channel
$\nu_i$. Therefore, we can compute the entries of the correlation matrix between each pair of frequency
channels

\begin{equation}
R_{jk}=\frac{C_{jk}}{C_{jj}^{1/2}C_{kk}^{1/2}},
\end{equation}
where the indices run from 1 to $N_f$. We diagonalise the correlation
matrix of the full data set with an eigenvalue decomposition and
obtain
  \begin{equation}
\transpose{P}RP=\Lambda \equiv \textrm{diag}\begin{Bmatrix}\lambda_1,...,\lambda_{N_f}\end{Bmatrix}, 
 \end{equation}
where the diagonal elements of the matrix $\Lambda$ are the
 eigenvalues $\lambda_j$ of the matrix $R$ and the matrix
 $P$ is an orthogonal matrix which contains the eigenvectors. The variance of each mode is given by the
 amplitude of the eigenvalues $\lambda_j$, so each eigenvalue measures
 the contribution of its corresponding eigenvector to the total sky
 variance.

 This method parameterises the foreground and noise components and
 produces independent eigenfunctions, which convert the spectral
 correlation into a number of largest variance modes. We pick the eigenvalues with the correlated components
 in frequency with the larger variances. 
 So, we build a matrix
 $P_{c}$, with only the corresponding eigenvectors and we use this matrix to decompose the data into
 eigenfunctions $\mathbf{\phi}$
   \begin{equation}\label{eq:pca_loss}
\mathbf{\phi}=\transpose{P_c}S.
 \end{equation}

The maps $S_{c}$ of the reconstructed foreground and
 $1/f$ noise are obtained by transforming back to the frequency
 space
   \begin{equation}
S_{c}=P_{c}\mathbf{\phi}. 
 \end{equation}
Finally, we find the maps of the reconstructed HI signal $S_{\textrm{HI}}$ by subtracting the input maps and the reconstructed foreground and $1/f$ noise
    \begin{equation}
S_{\textrm{HI}}=S-S_{c}. 
 \end{equation}

 \subsection{$1/f$ noise subtraction using PCA}\label{subsec:noise_sep}
First, we apply the PCA method to thermal and $1/f$ noise components only,
ignoring foregrounds for the moment. The frequency spectrum of the
correlated noise (1/$f$) is also expected to be smooth in frequency. Thus, one
can use the PCA method to remove the instrumental $1/f$ noise. We show
the result in Section~\ref{subsubsec:pcanoiserem} and we test the
robustness of this noise removal method with different models of the
$1/f$ noise in Section~\ref{subsubsec:noisefknee}.
 
 \subsubsection{PCA results}\label{subsubsec:pcanoiserem}
\begin{figure}
\begin{center}
\includegraphics[width=\columnwidth]{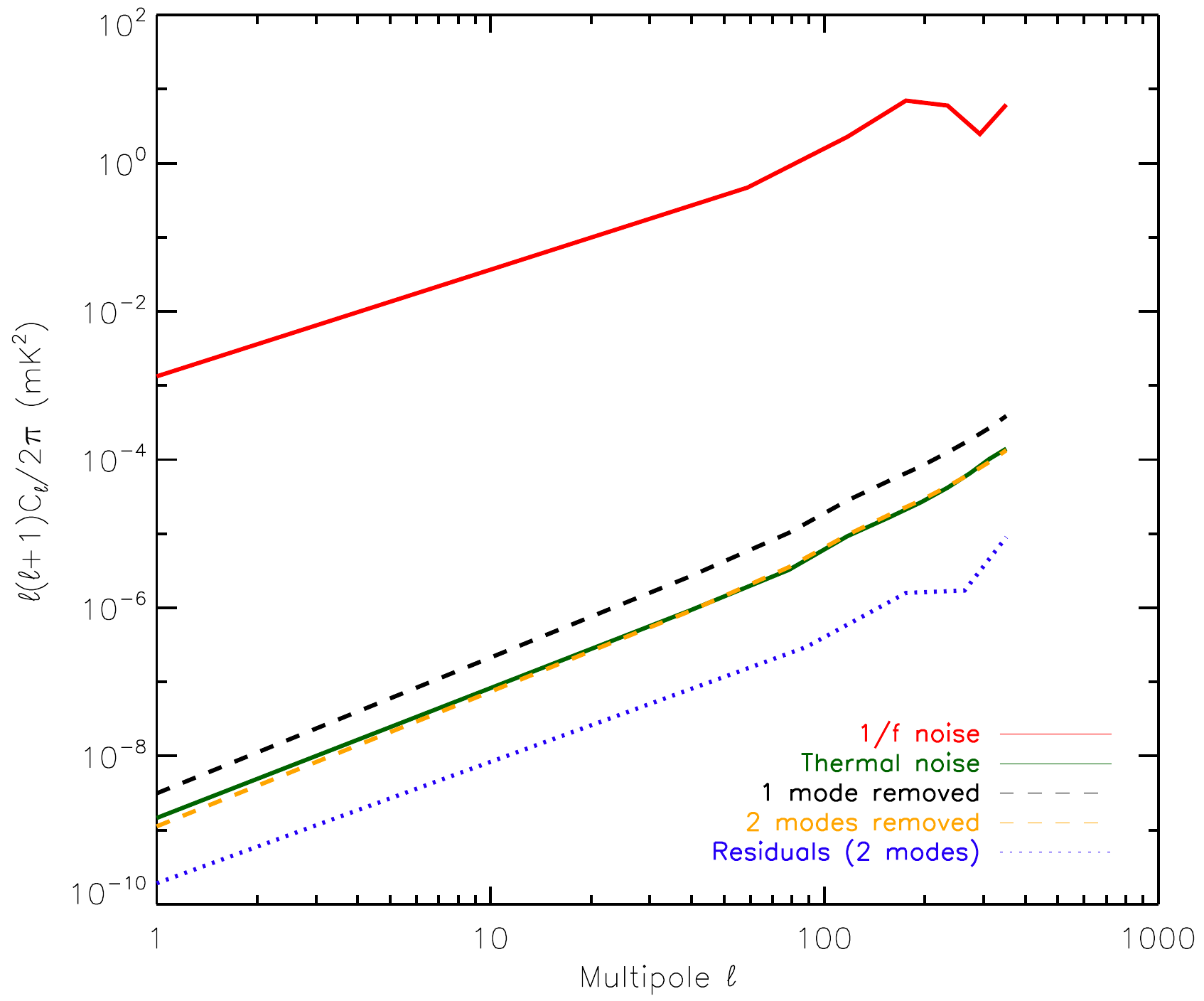}
\caption{Power spectra of the simulated noise maps after applying
principal component analysis. We show the input 1/$f$ noise (\textit{red solid line})
and the input thermal noise (\textit{green solid line}), the reconstructed noise after applying principal component
analysis with 1 mode removed (\textit{black dashed line}) and with 2 modes
removed (\textit{orange dashed line}). We display the power spectrum of the difference between the maps of the input thermal noise and of the reconstructed thermal noise after PCA with 2 modes removed (\textit{blue dotted line}). It is below the input
thermal noise at all scales.}
  \label{Fig:ps_noise}
\end{center}
\end{figure}

The instrumental noise is simulated as explained in
Section~\ref{subsec:instru_noise}, and we apply the PCA method to the
maps of the instrumental noise. The $1/f$ noise is computed with the
knee frequency $f_{\textrm{knee}}=1$\,mHz, which is thought to be achievable using balanced correlator receivers \citep{Jarosik2003, Bersanelli2010}. Note that a different scanning strategy than a drift-scan strategy can remove the $1/f$ noise in the map-making, but for a transit telescope, such as BINGO, we will rely on component separation and the smoothness of the frequency spectrum. In
the top panel of Fig.~\ref{Fig:ps_noise} we plot the power spectra of the noise maps
with different number of modes removed: 1, 2 and 3. The
spectra of the maps are computed to $\ell <$1000 using
\texttt{PolSpice} \citep{Szapudi2001,Chon2004,Challinor2005}. This code computes correlation functions and estimates the power spectra by integrating the resampled correlation function using Legendre-Gauss integration. The power spectra are corrected
for the effect of the cut sky and for the beam and pixel window
functions. In order to remove the ringing in the power spectra, we
apodize the correlation function with a Gaussian of width
15$^{\circ}$. Subtracting one mode does not remove the $1/f$ noise
sufficiently well, but the thermal noise level can be reached by removing 2 modes as displayed in Fig.~\ref{Fig:ps_noise}. This plot shows the residuals between the recovered thermal noise and the input thermal noise.  For
the case of two removed modes, the residual is significantly lower
than the input thermal noise at all scales. It shows that we
can recover the thermal noise model sufficiently well using principal
component analysis by subtracting at least 2 principal modes. 

\subsubsection{Results with different noise models}\label{subsubsec:noisefknee}
 
In order to test the efficiency of the PCA method, we compute the
instrumental noise with different values of the knee frequency
$f_{\textrm{knee}}$ between 1\,mHz and 10\,Hz. Note that a knee
frequency of 1\,mHz might be expected for a pseudo-correlation receiver and
that 10\,Hz is a worst-case scenario for a single channel radiometer. We
find that the residual noise after removing 2 principal modes is independent of the input knee frequency and, hence, the PCA method is robust. We emphasise that we have assumed a flat frequency spectrum for
the $1/f$ noise (i.e. it affects all frequency channels equally). The efficiency of the noise cleaning method depends on this assumption and on assuming a perfect calibration.
We will quantify the success of this method as
the function of the smoothness of the $1/f$ noise in
Section~\ref{sec:smoothness}.
 
\subsection{Inclusion of foreground emission}\label{subsec:res_sep_meth}
In this Section, we show the results for the detection of the HI
signal from the total intensity maps in the presence of instrumental
noise and foreground emission, using the parametric fitting and principal component analysis methods. We compare both methods
in Section~\ref{subsubsec:pf_res} and we focus on the results of the principal component analysis
method in Section~\ref{subsubsec:pca_res}. We look at the residuals of
the reconstructed cosmological signal in Section~\ref{subsubsec:ps_res}.

\subsubsection{Parametric fitting results}\label{subsubsec:pf_res}
\begin{figure}
 \begin{center}
\includegraphics[width=\columnwidth]{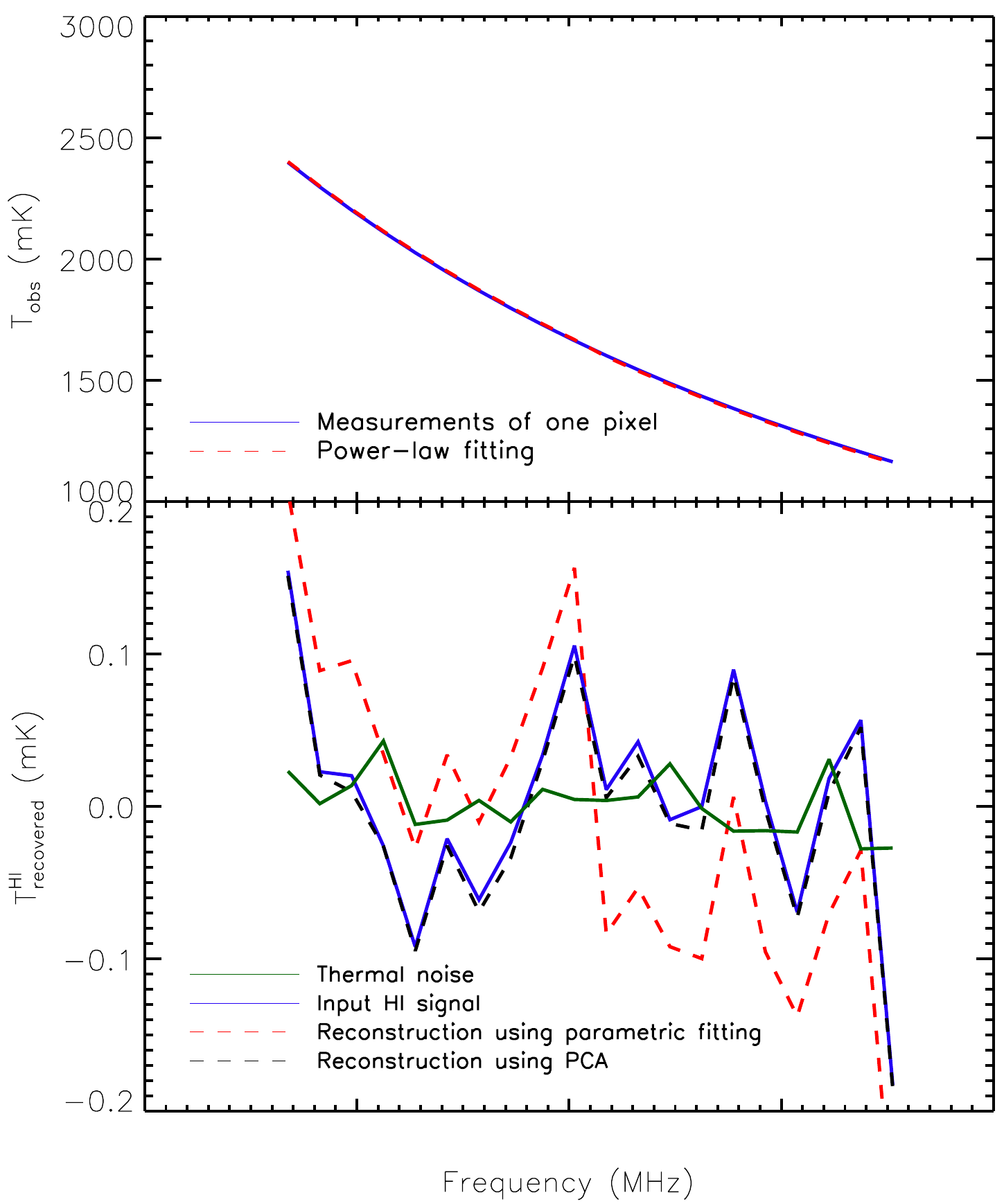}
\caption{\textit{Top} panel: brightness temperature of one particular
line-of-sight as a function of frequency composed of sky emissions and instrumental noise (\textit{blue solid line}), averaged over 20 realisations and the fit based on a parametric fitting (\textit{red dashed line}). \textit{Bottom}
panel: the recovered HI signal from parametric fitting (\textit{red dashed line}) and from principal component analysis (\textit{black dashed line}). We show the input HI (\textit{blue solid line}) and the thermal noise (\textit{green solid line}).  }
  \label{fig:plotPLF}
\end{center}
\end{figure}

Here we present the results of the parametric fitting method. The
extraction of the HI signal is done using only the frequency
information. The fit is made for each pixel. In Fig.~\ref{fig:plotPLF}, we show the measurements as a
function of frequency, for a random line-of-sight, with the
synchrotron model 3 quantified in Table \ref{tab:synchrmodel} and a background of unresolved point sources ($S< 100$\,mJy) as explained in Section~\ref{subsubsec:ps}. The result is averaged over 20 realisations of the
instrumental noise (thermal and $1/f$ noise). The top
panel represents the simulated measurements and the reconstructed
foreground emission with parametric fitting, highlighting the
smooth component of the foreground. The bottom panel represents the
recovered cosmological signal with parametric fitting and principal
component analysis after removing 7 modes compared to the input
one. It shows that the parametric fitting, while superficially in agreement with the input signal, does not provide an accurate
fit to the signal of interest compared to the PCA method.

\subsubsection{Results of PCA applied to sky maps}\label{subsubsec:pca_res}

Fig.~\ref{fig:plotPLF} shows that PCA induces a small offset in the reconstructed HI signal: some
cosmological signal leaks into the reconstructed foreground and noise
components. However, with this method, the HI signal is well recovered
with a relative error of $\sim$7\%.

\begin{figure}
\begin{center}
\includegraphics[width=\columnwidth]{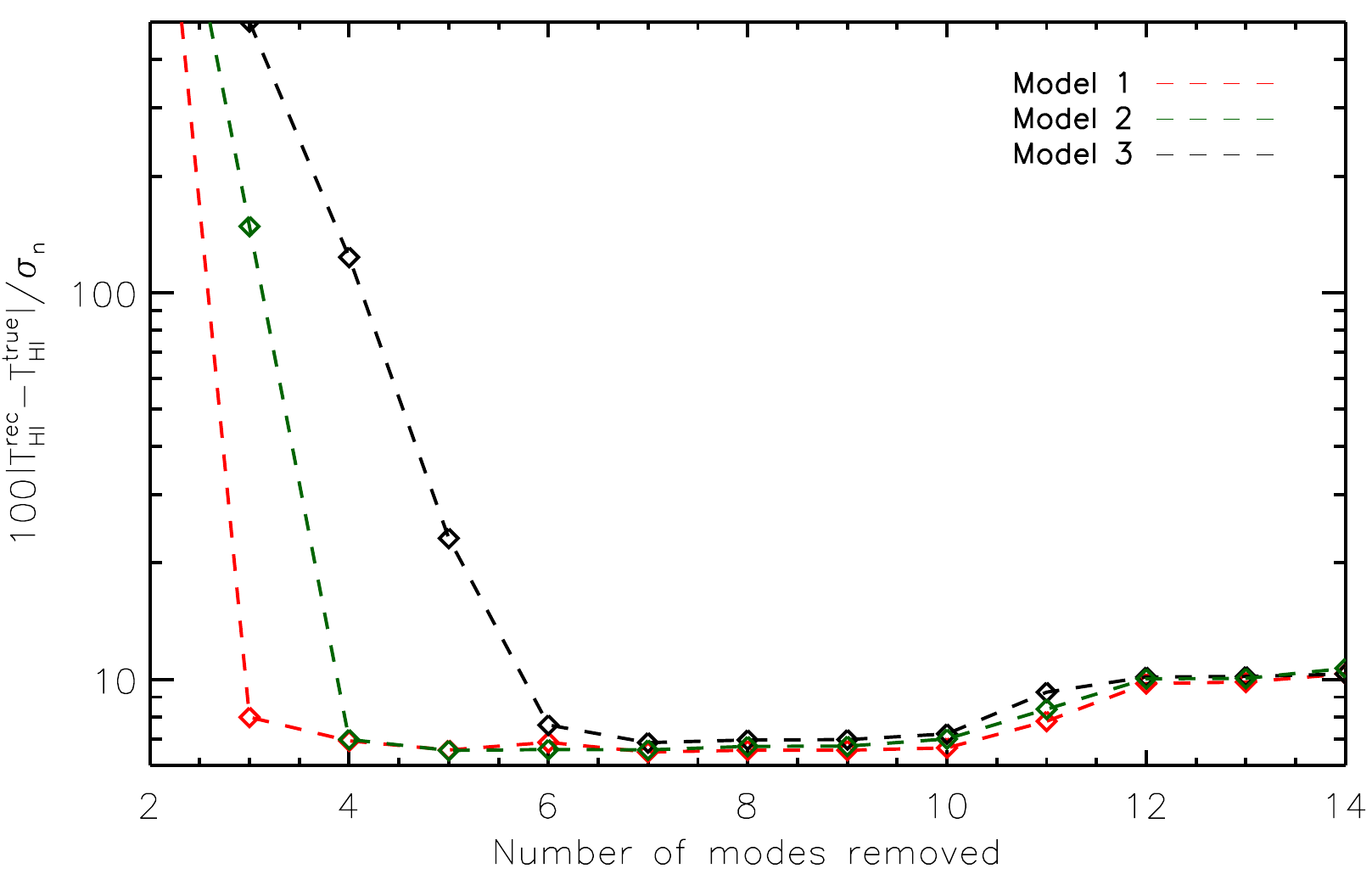}
\caption{The percentage relative error of the thermal noise as a function
of the number of modes removed. We plot the error for a
foreground model with a background of unresolved point sources and
Galactic synchrotron emission with a constant index $\beta=-2.8$ (\textit{red dashed line}), with the introduction of the random spatial
distribution of the index $\beta$ across the sky (\textit{green dashed line}), and with correlated spatial and spectral variations of the index
$\beta$ plus a background of unresolved point sources (\textit{black dashed line}). These models correspond respectively to models 1, 2 and 3 explained in Table \ref{tab:synchrmodel}. The
relative error increases significatively above 10 modes due to the
leakage of the cosmological signal into the reconstructed foreground and
correlated noise components.}
  \label{fig:residu_pca}
\end{center}
\end{figure}

Here we will quantify the impact of the foreground residuals on the reconstructed
HI signal after applying the PCA method to the simulated sky maps.  In
order to check the effectiveness of the foreground removal method,
we define the relative error by
\begin{equation}
 |T_{{\textrm{HI}}}(\nu)-\hat{T}_{{\textrm{HI}}}(\nu)|/ \sigma_\textrm{n}, 
 \end{equation}
where $T_{{\textrm{HI}}}(\nu)$ is the true HI signal at frequency $\nu$, $\hat{T}_{{\textrm{HI}}}(\nu)$ the recovered signal,
and $\sigma_\textrm{n}$ the standard deviation of the thermal noise.
Fig.~\ref{fig:residu_pca} represents the relative error as a function
of the number of subtracted modes for a simulation with HI signal,
$1/f$ and thermal noise and different models of foreground. We apply the PCA technique to each of the foreground models discussed in Section~\ref{subsubsec:synch}.

Since the foreground and the $1/f$ noise spectra do not contain sharp
features, we can expect that they are well described by a small number
of eigenvectors, so the eigenvalues are much larger for the first few
principal components. This implies that a small number of components contains
almost all of the foreground emission and the $1/f$ noise.  Fig.~\ref{fig:residu_pca} shows a fall-off of the amplitude of the eigenvalues with an increase
in the value of the number of principal components. This steep drop
means that the spectra are dominated by relatively few components, which are related to the foreground and smooth instrumental
contamination. Furthermore, this figure shows that the foreground model has an
impact on the extraction of the HI signal as the most complex foreground
model (model 3) requires more modes to be removed in order to subtract the same level of foreground contamination. With the same noise model, the first
foreground model requires the removal of at least 3 principal modes, the
second model, 4 modes and the third model, 7 modes.

We notice that, when a larger number of modes is removed beyond
a certain threshold, the relative error associated with each foreground
model begins to increase. This can be understood in terms of the
component separation method inducing a leakage of the cosmological
signal in the foreground eigenvectors. This becomes even more
important when we remove a larger number of principal modes from the
initial maps. For the most realistic foreground model, model 3, the
cosmological signal is well recovered with a relative error of
$\sim$7\% after subtracting 7 modes and the percentage error
increases to $\sim$9.5\% with 11 modes removed.

To evaluate the performance of the PCA method, we calculate the amount of
leakage of the HI signal into the subtracted modes using
Eq.~\ref{eq:pca_loss}. The relation between the maps of the HI signal
at each frequency channel $S_{\textrm{HI}}$ and the orthogonal matrix
$P$ which contains the principal modes of the maps is given
by the eigenfunctions $\mathbf{\phi}$
  \begin{equation}
\mathbf{\phi}=\transpose{P}S_{\textrm{HI}}.
 \end{equation}
Finally, we obtain the maps of the leakage of the HI signal
$S_{\textrm{HI}}^{'}$ using the orthogonal matrix $P_{c}$,
which contains only the eigenvectors removed from the initial maps
  \begin{equation}
S_{\textrm{HI}}^{'}=P_{c}\mathbf{\phi}.
 \end{equation}
Removing 7 principal modes with the PCA method enables one to recover the HI
signal. However, this  induces a loss of the cosmological signal
of $\sim$5\%. In order to determine the foreground component that has the most
impact on the foreground separation, we generate maps of the sky
emission with only synchrotron emission and HI signal, and maps with a
background of unresolved point sources and HI signal. The maps with
synchrotron emission require the removal of 3 modes in order to
extract the signal of interest, and we obtain a leakage of the HI
signal in the removed principal modes of $\sim$2.4\%. To recover the
HI signal from maps with only a background of unresolved point sources, the
subtraction of 1 mode is required and leads to a leakage of
$\sim$1.6\% of the cosmological signal. Subtracting too many principal
modes induces a significative loss of the cosmological signal, while when an insufficient number of modes is removed, the recovered HI signal will
be still affected by the foreground contamination.

Up to now we have assumed a perfect smoothness of the
foreground spectra. This assumption could be broken in the presence of
instrumental systematic effects such as not sufficient 
knowledge of the beams, imperfect polarisation purity and
mis-calibration will affect the results of the component separation
methods, adding additional uncertainties. To measure the BAO wiggles from the
HI power spectrum, we require that the statistical error dominates the
errors from foreground cleaning methods and calibration. For the
present analysis, we neglect calibration systematics, postponing
discussion of the required calibration accuracy needs to be done to a future
paper. We are however aware that accurate calibration of bandpasses
and beam polar diagrams is essential. Quantifying calibration requirements
will be done with an end-to-end pipeline. We are also
investigating a more complex foreground cleaning method, which uses combined spatial and spectral filtering techniques based on the
expertise from the CMB (e.g. \citealp{Leach2008, Remazeilles2011} and Olivari et al. in prep.).

\subsubsection{Maps of the reconstructed HI signal and power spectra}\label{subsubsec:ps_res}
\begin{figure*}
\centering
{

\includegraphics[width=12.5cm,height=0.5cm]{signal.png}
}
\quad
{
\subfigure[Input HI signal]{
\includegraphics[width=5.5cm]{bar.pdf}}
}
\quad

{
\includegraphics[width=12.5cm,height=0.5cm]{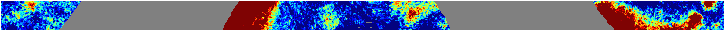}
}
\quad
{
\subfigure[PCA 1 mode removed]{
\includegraphics[width=5.7cm]{bar.pdf}
}}
\quad
{
\includegraphics[width=12.5cm,height=0.5cm]{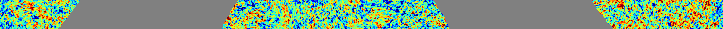}
}
\quad
{
\subfigure[PCA 3 modes removed]{
\includegraphics[width=5.7cm]{bar.pdf}}
}
\quad
{
\includegraphics[width=12.5cm,height=0.5cm]{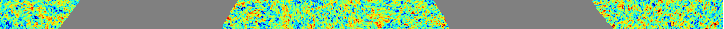}
}
\quad
{\subfigure[PCA 7 modes removed]{
\includegraphics[width=5.5cm]{bar.pdf}}
}
\quad
{
\includegraphics[width=12.5cm,height=0.5cm]{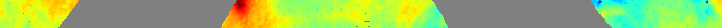}
}
\quad
{\subfigure[Parametric fitting]{
\includegraphics[width=5.5cm]{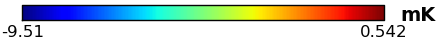}}
}
\caption{Five versions of a strip in declination of the HI signal. The input cosmological signal is shown in the first strip and the reconstructed signal from principal component analysis in strips 2, 3 and 4 (after 1, 3 and 7 modes removed respectively) and from parametric fitting in strip 5. Notice the different colour bar scales. }
  \label{fig:fitting_res}
\end{figure*}

We perform the foreground cleaning using the PCA method with 7 modes
subtracted and the parametric fitting. In Fig.~\ref{fig:fitting_res},
we display the maps of the recovered HI signal for the foreground
model 3. The simulations contain instrumental noise ($1/f$ and thermal
noise). To compare with the results of the cleaning methods, we plot
in the first strip the true HI signal. The second, third and fourth
strips show the reconstructed HI signal after 1, 3 and 7 principal modes are removed
respectively. One can notice the similarity between the reconstructed HI signal shown in the fourth strip and the strip of the true HI signal. These two maps
show that we can extract the cosmological signal from a highly
contaminated map. The fifth strip represents the recovered HI signal
after applying parametric fitting. Noting that the temperature scales
are different, we observe significant differences between the input HI
strip and that recovered with parametric fitting. It is clear that parametric fitting
does not provide a sufficiently good fit to foreground and noise components.

\begin{figure}
\begin{center}
\includegraphics[width=\columnwidth]{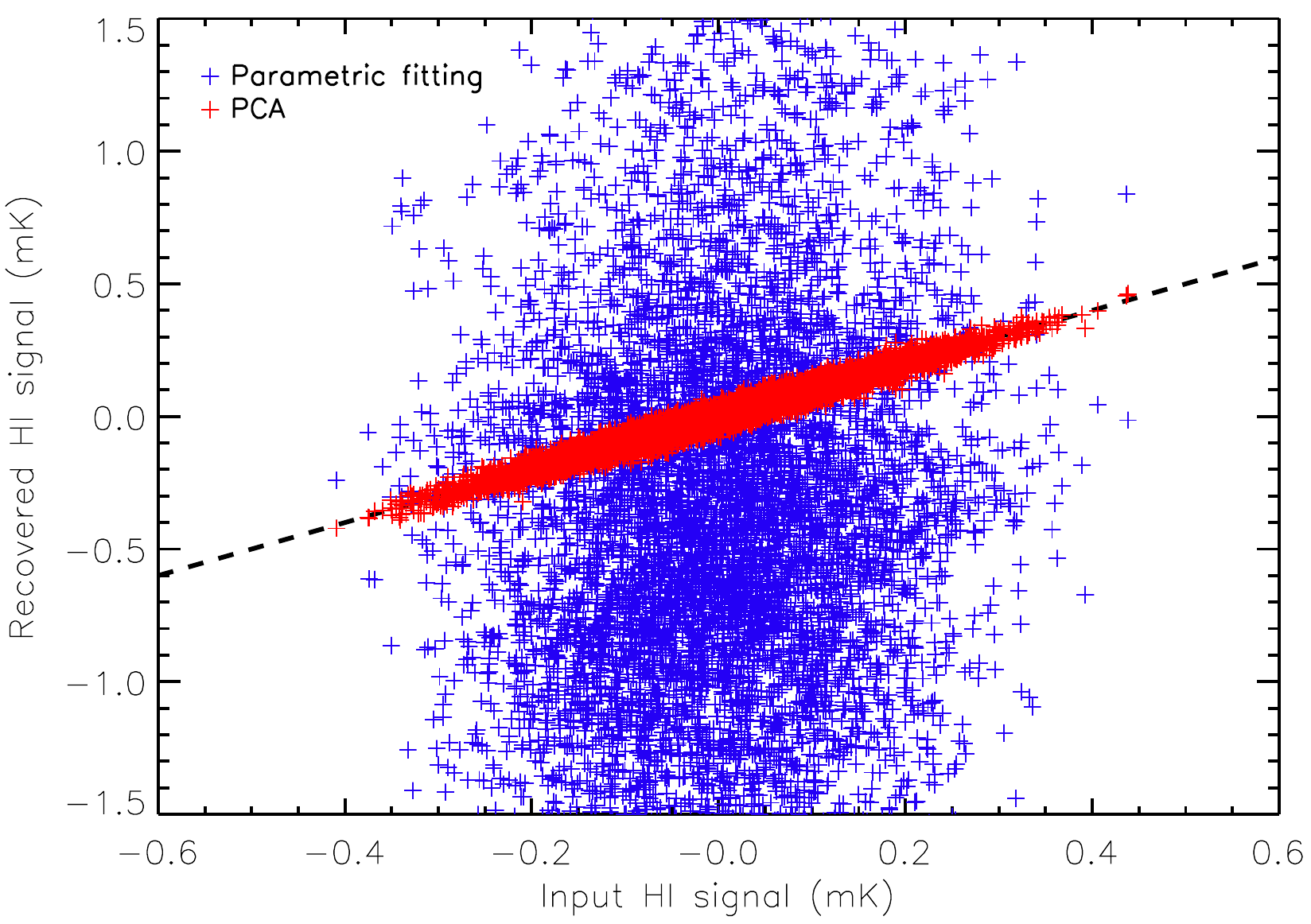}
\caption{Scatter plot of the recovered HI signal from parametric fitting (\textit{blue markers}) and from principal component analysis (\textit{red markers}) as a function of the input HI signal. The parametric fitting is more noisy than the principal component analysis and also appears to be biased. The black dashed line represents the perfect correlation between the recovered HI signal and the true signal.}
\label{fig:tt}
\end{center}
\end{figure}

It is evident that parametric fitting does not remove the foreground completely, thus
resulting in foreground leaking into the reconstructed
cosmological signal. To highlight the comparison between parametric fitting and PCA, 
we show in Fig.~\ref{fig:tt} the dispersion, pixel by pixel, between the recovered HI signal as
a function of the input HI signal, obtained from parametric fitting and
from PCA. This plot shows that parametric fitting is much less
effective than the PCA method and induces a bias in the reconstructed
HI signal. 

\begin{figure}
\begin{center}
\includegraphics[width=\columnwidth]{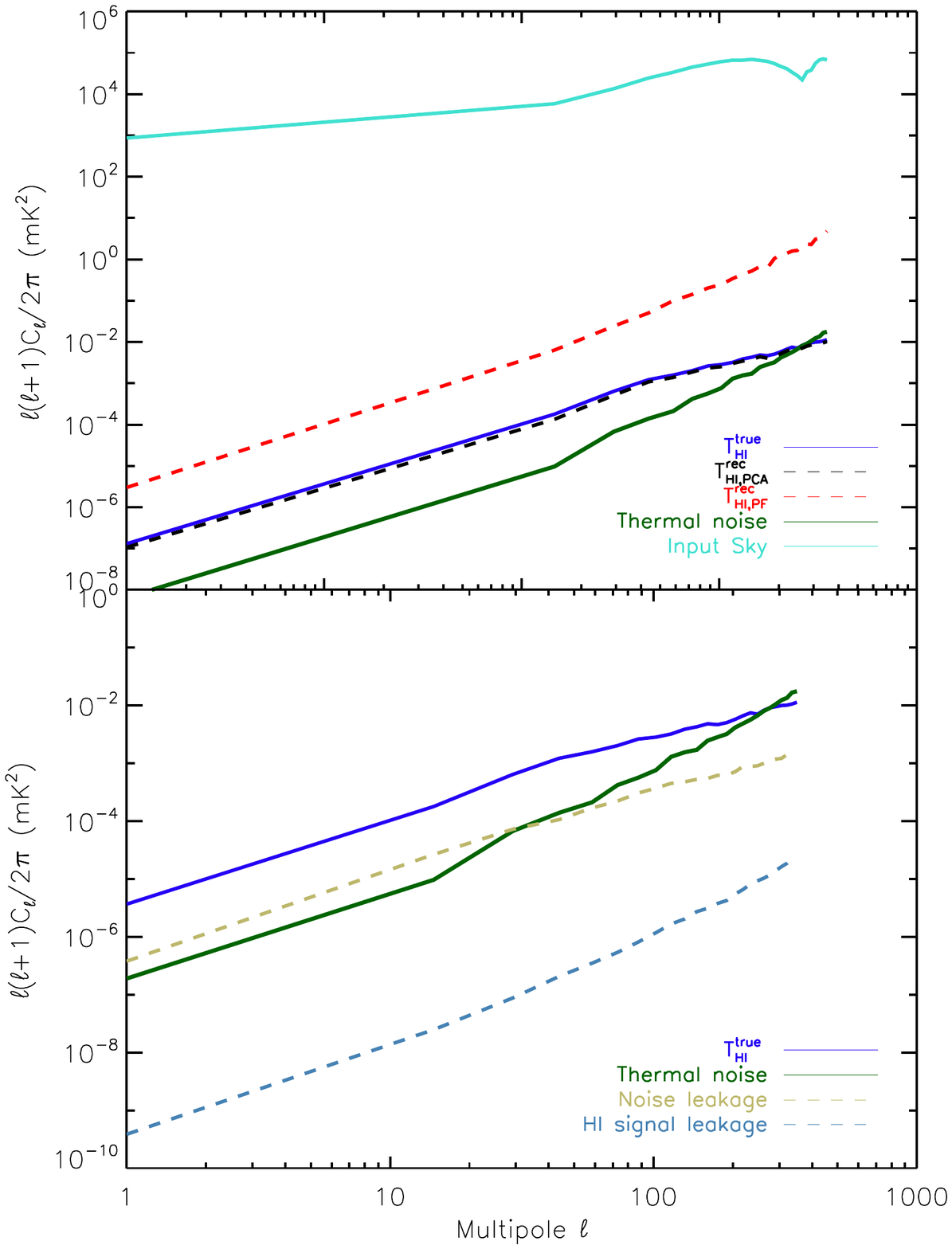}
\caption{\textit{Top} panel: the power spectra of the simulated maps after applying foreground cleaning. This plot shows the simulated HI signal (\textit{blue solid line}), the thermal noise (\textit{green solid line}) and the sky emission (\textit{cyan solid line}). We plot the results of the foreground cleaning from principal component analysis (\textit{black dashed line}) and from parametric fitting (\textit{red dashed line}). \textit{Bottom} panel: the power spectra of the leakage of the HI signal (\textit{dotted light blue line}) and of the noise (\textit{gold dashed line}) after applying PCA. The principal component  analysis method clearly makes it possible to extract the HI signal.}
  \label{fig:ps_foreground}
\end{center}
\end{figure}

We can quantify the leakage of the thermal noise and of the
cosmological signal into the reconstructed foreground and noise
components by calculating their power spectra and comparing them to
their input power spectra. In the bottom panel of Fig.~\ref{fig:ps_foreground}, we display
the power spectra of the true and the recovered HI signal after the
removal of 7 modes. This figure shows that both parametric fitting
and PCA methods remove several orders of magnitude of foreground
contamination and 1/$f$ noise, but PCA gives lower residuals than the parametric fitting method. We plot the cosmological signal leakage into the foreground and $1/f$ noise reconstruction in the bottom panel. The power
spectra of the thermal noise leakage and the HI signal leakage are
lower than the input HI signal at all scales, thus, with PCA, it is
feasible to extract the HI signal from a highly contaminated
foreground map.

\section{Requirements on foreground and instrumental noise frequency spectra}\label{sec:smoothness}
For almost all component separation techniques developed for intensity
mapping data analysis, the efficiency of these methods depends on the
spectral smoothness of the foreground and that of the $1/f$ noise and the bandpass calibration. We
are confident that the foregrounds have sufficiently smooth spectra in
frequency. This characteristic enables us to remove the components
correlated in frequency and, therefore, to recover the HI signal with
the instrumental white noise. However, some structure in the receiver
band-passes is inevitable caused by amongst other things, standing
waves, frequency variations in the receiver gain temperature and
spectrally dependent beam patterns. The effect of these can all be
reduced by careful calibration but it is important to know how good
this calibration has to be. In this section we define some
requirements on the smoothness of the instrumental bandpass. In
the following, for the foreground emission, we consider only the
brightest component; i.e. the Galactic synchrotron emission.

To quantify the effect of a non-smooth bandpass, we add to
the measured frequency spectrum a sinusoidal wave defined
by
\begin{equation}
\phi(\nu)=A \, \text{\textnormal{sin}} \left( \frac{ \pi \nu}{\Delta \nu} \right),
\label{eq:sim}
\end{equation}
where $A$ is the amplitude, $\nu$ the frequency
of observation and $\Delta \nu$ the wavelength. We explore a range of $A$ between 1 and 150\,mK and a range of
$\Delta \nu$ between 1 to 300\,MHz.

\begin{figure}
\begin{center}
{\includegraphics[width=\columnwidth]{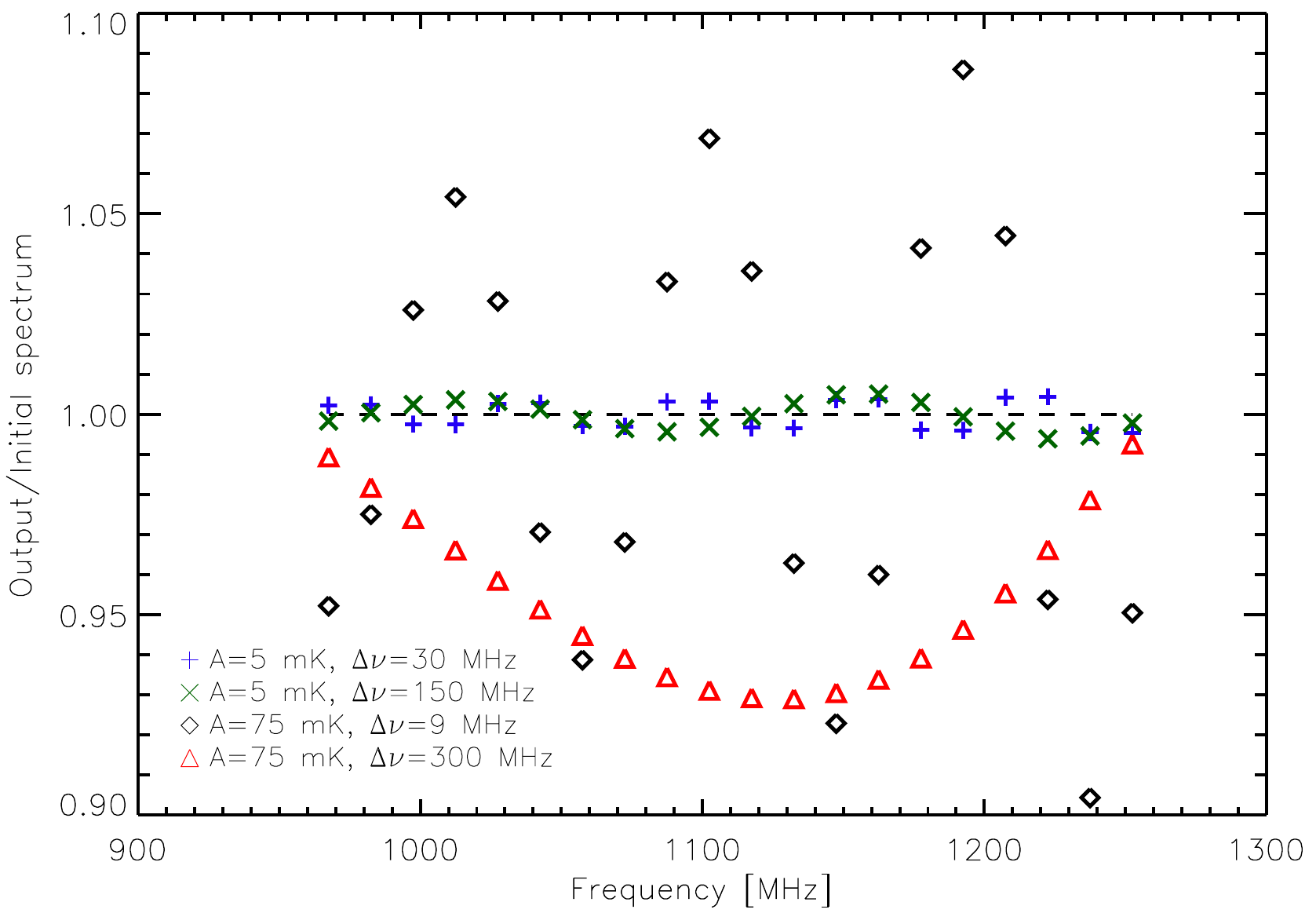}
\caption{The corrupted spectrum divided by the
original, undistorted, spectrum for different versions of the sinusoidal
wave. A small value of $\Delta \nu$ induces a spectrum that
fluctuates in frequency in a similar way to random noise, while a
large value of $\Delta \nu$ leads to a curvature of the frequency
spectrum.  }
  \label{fig:ex_req}
}\end{center}
\end{figure}

We show the modified spectra for different values of $A$
and $\Delta \nu$ in Fig.~\ref{fig:ex_req}. To highlight the impact of
the addition of the sinusoidal wave, we divide the resulting 
spectrum by the original one. We see curvature and/or
oscillations in the resulting spectra. A higher value of
$\Delta \nu$ leads to a curvature of the spectrum, similar
to a standing wave, whereas a smaller value induces a sinusoidal wave
that behaves in a similar way to noise, when $\Delta \nu$ is smaller
than the frequency channel width.

\begin{figure}
\includegraphics[width=\columnwidth]{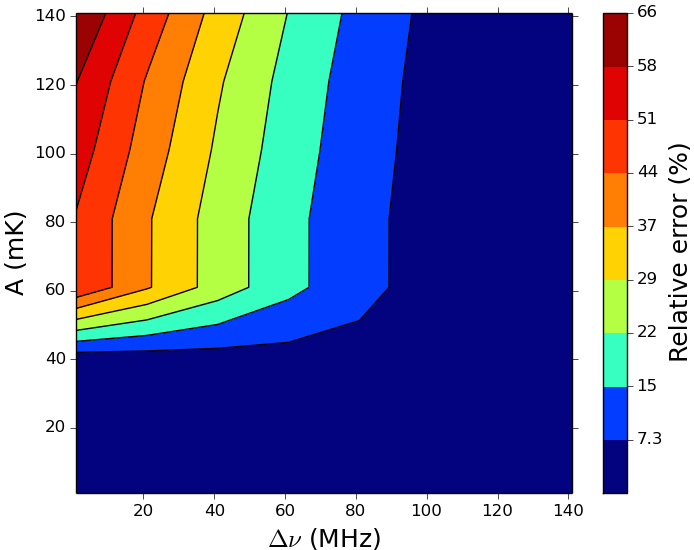}
\caption{Relative error as a function of the amplitude $A$ and the period $\Delta \nu$ of the sinusoidal wave after foreground and noise subtraction after applying principal component analysis (6 modes removed). The values $A$ and $\Delta \nu$ of the sinusoidal waves are indicated on the axes of the plot. The colour bar gives the percentage error relative to the noise.}
 \label{fig:contour_sky}
\end{figure}

In what follows, we simulate the maps generated by the instrument with the same model of the Galactic
synchrotron emission, HI signal and instrumental noise (thermal and
1/$f$ noise) and we add Eq.~\ref{eq:sim} the sinusoidal wave to the frequency spectrum of the generated data. In order to extract the signal of interest, we apply the principal component analysis to
the maps. In Fig.~\ref{fig:contour_sky}, we plot the relative error of the
recovered HI signal as a function of the amplitude $A$ and $\Delta
\nu$ after applying PCA with 6 modes subtracted. The colour bar
represents the amplitude of the relative error between the recovered
HI signal and the true signal. The smoothness of the frequency
spectrum, i.e. the value of $\Delta \nu$, has a significative impact
on the efficiency of the cleaning methods. The relative error
increases with a small value of $\Delta \nu$, which corresponds to a
sinusoidal wave with a period shorter than the frequency channel
width. In order not to be affected by the variation of the bandpass,
the value of $\Delta \nu$ has to be lower than 100\,MHz and the
amplitude $A$ has to be below 45\,mK. With the values $A<40$\,mK and
$\Delta \nu<100$\,MHz, we find a relative error $<7.3$\% after 6 modes are subtracted with the PCA. In absolute terms, after subtracting 6 principal
modes, we obtain residuals lower than 0.1\,mK, which means that the HI
signal can be detected.

\begin{figure}
\includegraphics[width=\columnwidth]{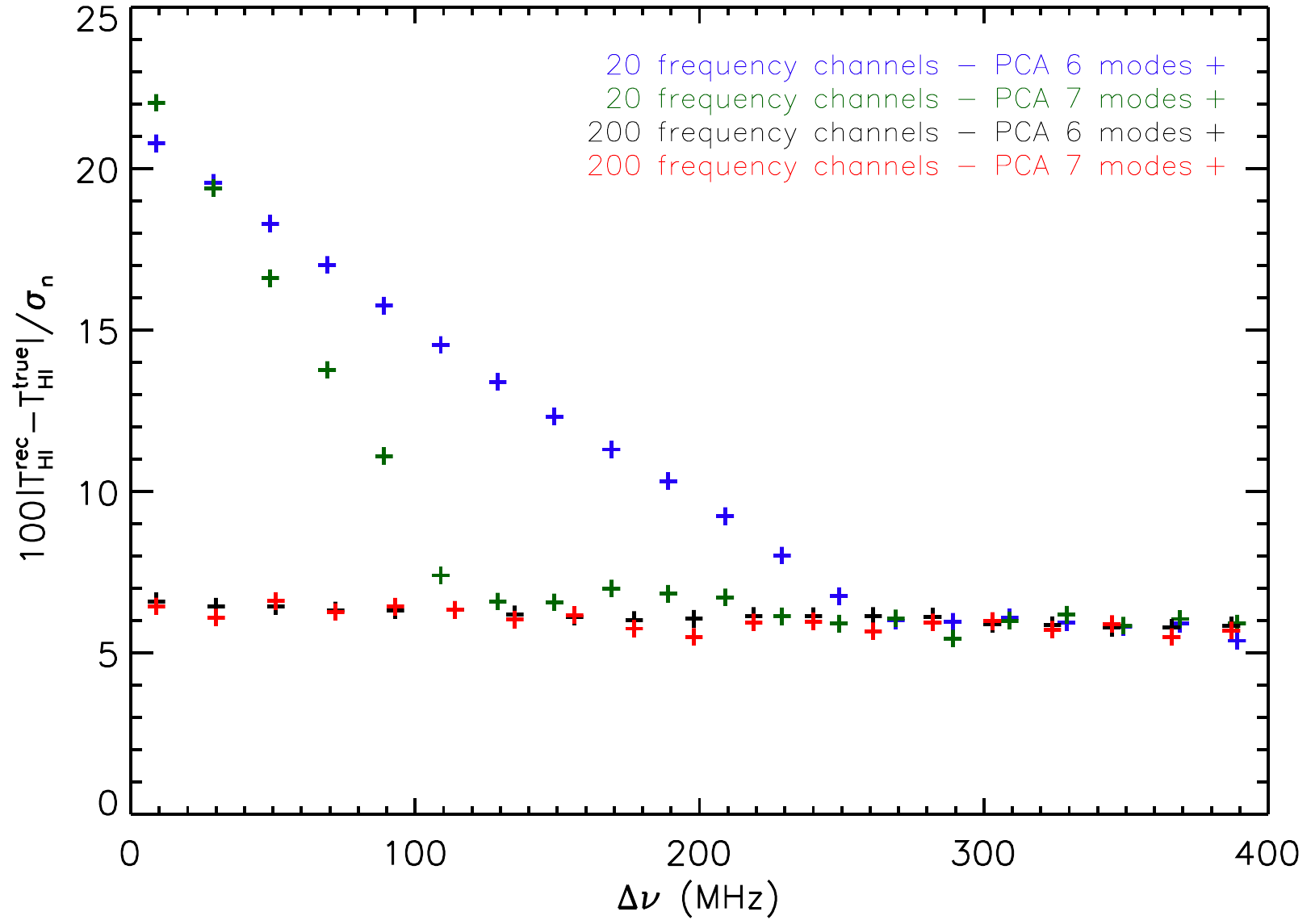}
\caption{The percentage relative error as a function of the period $\Delta \nu$ of the sinusoidal wave after foreground removal. We show the relative errors for 20 frequency channels with 6 modes removed (\textit{blue markers}), for 20 frequency channels with 7 modes removed (\textit{green markers}) and for 200 frequency channels with  6 and 7 modes removed (\textit{red and black markers}) respectively.}
 \label{fig:nbchannel}
\end{figure}

Finally, we perform simulations varying the number of frequency
channels used to perform the PCA. We consider 20 frequency channels
(15\,MHz channel bandwidth) and 200 frequency channels (1.5\,MHz channel
width) and we test different values of $\Delta \nu $. We choose the
amplitude of the sinusoidal wave to $A=120$\,mK. Fig.~\ref{fig:nbchannel} shows the relative error between the
recovered HI signal and the input signal as a function of the
smoothness of the frequency spectrum $\Delta \nu$ after removing 6 and
7 principal modes. We find that the PCA method does better with a
larger number of channels. The relative error is $<7$\% for 6 removed modes when we have 200 frequency channels for all values of $\Delta
\nu$ between 1 to 400\,MHz. The reason for the improvement when more
channels are added can be understood by the fact that the frequency
band is better sampled. Thus, as long as we have a frequency spectrum
with slow oscillations, or enough frequency channels to sample the
spectrum with sufficient accuracy, the smoothness of the bandpass
does not constitute an issue for the foreground and the $1/f$ noise
cleaning methods. An amplitude around 40\,mK requires the bandpass
to be calibrated to an accuracy of better than 1 part in 1000. However, one would expect to calibrate at least every day so we will only require a dynamic range of 1 part in 50.

\section{Conclusion}

In Section~\ref{sec:simu}, we have simulated the kind of data which
would be produced in a single-dish intensity mapping experiment like
BINGO. We have adopted a sky model with Galactic synchrotron emission
and a background of point sources ($<10$\,mJy). We have added both white and $1/f$ noise. This
simulation can be generalised to any intensity mapping experiment with
a drift scan strategy. This work has yielded several results
concerning the challenge of any future single-dish experiment in terms
of foreground, atmospheric and instrumental noise.

We have made some estimates of the total amplitude of the
atmospheric noise and of its fluctuations arising from turbulences in
emission of the water vapour. At 1\,GHz, we have found that the r.m.s. noise level
induced by these fluctuations is $\sim 0.01$\,mK below the instrumental
noise level for the BINGO experiment. Thus, for an observing frequency around 1\,GHz, the
atmospheric contamination should not constitute a challenge for a single-dish experiment. 

We have investigated the problem of cleaning foreground contamination
in order to recover the HI signal. We have focused on simple methods
based on the spectral information contained in the signal. We have
considered two different foreground cleaning methods, a non-blind
method with parametric fitting and a blind method with PCA. This last method does not require any assumption
on the physics of the foreground emissions. We have shown that the
parametric fitting method does not provide an accurate fit to the data
for a simulation with instrumental noise and foreground emissions. In
contrast, by subtracting 7 principal modes, with a realistic
foreground model, PCA enables to reach the
HI signal level. Thus, on simulated data, the application of the PCA method shows that it is feasible to
extract the cosmological signal across a wide range of multipoles and
redshifts.

However, we have found that PCA induces a
small offset in the reconstructed the HI signal of $\sim$5\%. This result
is confirmed by other recent works \citep{Alonso2015}, which have
shown that it is possible to recover successfully HI signal from a highly
contaminated foreground map with a blind method but that the
foreground removal induces a bias in the HI power spectrum. In spite
of this HI leakage into the removed contamination, a modification of
the global shape of the power spectrum should not modify the positions of the BAOs
wiggles \citep{Wolz2014} and therefore will not have significant impact on our ability to measure the
BAO scale. If the PCA method is chosen, we can correct for this small bias by calibrating it with simulations. We are also investigating a more complex cleaning method, which utilises both the spectral and spatial information (Olivari et al., in prep.), and therefore should be more accurate.

Given the assumption that the frequency spectrum of the $1/f$ noise is
flat, or at least only has slow variations across the band, we have
shown that PCA can cope with 1/$f$ noise with relatively high knee
frequencies. It is easily removed from the noise maps by subtracting 2
principal modes and the 1$/f$ noise contamination is dampened down to
the level of the thermal noise. This result implies that it might be
possible to simplify the design of an experiment, such as BINGO,
by removing the correlation receivers. In subsequent work, we plan to
investigate more complex 1/$f$ noise models.

The effectiveness of the blind cleaning methods depends on the
spectral smoothness of the foregrounds and the instrumental $1/f$
noise. In order to challenge our cleaning methods, we have considered
instrumentally distorted frequency spectra of the components we want
to remove. We have shown that a sinusoidal distortion of the frequency
spectrum can induce higher residuals on the maps. However, for a
sinusoidal wave with an amplitude $A<45$\,mK and period $\Delta \nu<100$\,MHz, after subtracting six principal modes from the initial maps,
the residuals of the noise and the sky emission are below 0.1\,mK,
which could enable the successful extraction of the HI signal. Still, accurate calibration will be critical for the success of measuring Baryon Acoustic Oscillations and probing the expansion history of the Universe. How good this calibration needs to be will be
investigated in a future paper. The results of this paper show that
the principal component analysis method is a promising tool for the
extraction of the cosmological signal in any future HI intensity
mapping experiment.

\section{Acknowledgements}
MABS, CD, MR and YZM acknowledge support from an ERC Starting Grant (no. 307209). We acknowledge use of the \texttt{HEALPix} package \citep{Gorski2005} and the python tools, PyOperators and PySimulators \citep{Chanial2012}. We would like to thank the BINGO collaboration for providing the basic concept for which this work was based.

\bibliography{BINGO_bigotsazy_arxiv}

\begin{thebibliography}{70}
\expandafter\ifx\csname natexlab\endcsname\relax\def\natexlab#1{#1}\fi

\bibitem[{{Alonso} {et~al}\mbox{.}(2015){Alonso}, {Bull}, {Ferreira}, \&
  {Santos}}]{Alonso2015}
{Alonso} D., {Bull} P., {Ferreira} P.~G., {Santos} M.~G., 2015, \mnras, 447,
  400

\bibitem[{{Ansari} {et~al}\mbox{.}(2012){Ansari}, {Campagne}, {Colom}, {Le
  Goff}, {Magneville}, {Martin}, {Moniez}, {Rich}, \& {Y{\`e}che}}]{Ansari2012}
{Ansari} R. {et~al.}, 2012, \aap, 540, A129

\bibitem[{{Battye} {et~al}\mbox{.}(2013){Battye}, {Browne}, {Dickinson},
  {Heron}, {Maffei}, \& {Pourtsidou}}]{Battye2013}
{Battye} R.~A., {Browne} I.~W.~A., {Dickinson} C., {Heron} G., {Maffei} B.,
  {Pourtsidou} A., 2013, \mnras, 434, 1239

\bibitem[{{Battye}, {Davies} \& {Weller}(2004){Battye}, {Davies}, \&
  {Weller}}]{Battye2004}
{Battye} R.~A., {Davies} R.~D., {Weller} J., 2004, \mnras, 355, 1339

\bibitem[{{Bersanelli} {et~al}\mbox{.}(2010){Bersanelli}, {Mandolesi},
  {Butler}, {Mennella}, {Villa}, {Aja}, {Artal}, {Artina}, {Baccigalupi},
  {Balasini}, {Baldan}, {Banday}, {Bastia}, {Battaglia}, {Bernardino},
  {Blackhurst}, {Boschini}, {Burigana}, {Cafagna}, {Cappellini}, {Cavaliere},
  {Colombo}, {Crone}, {Cuttaia}, {D'Arcangelo}, {Danese}, {Davies}, {Davis},
  {de Angelis}, {de Gasperis}, {de La Fuente}, {de Rosa}, {de Zotti},
  {Falvella}, {Ferrari}, {Ferretti}, {Figini}, {Fogliani}, {Franceschet},
  {Franceschi}, {Gaier}, {Garavaglia}, {Gomez}, {Gorski}, {Gregorio}, {Guzzi},
  {Herreros}, {Hildebrandt}, {Hoyland}, {Hughes}, {Janssen}, {Jukkala},
  {Kettle}, {Kilpi{\"a}}, {Laaninen}, {Lapolla}, {Lawrence}, {Lawson}, {Leahy},
  {Leonardi}, {Leutenegger}, {Levin}, {Lilje}, {Lowe}, {Lubin}, {Maino},
  {Malaspina}, {Maris}, {Marti-Canales}, {Martinez-Gonzalez}, {Mediavilla},
  {Meinhold}, {Miccolis}, {Morgante}, {Natoli}, {Nesti}, {Pagan}, {Paine},
  {Partridge}, {Pascual}, {Pasian}, {Pearson}, {Pecora}, {Perrotta},
  {Platania}, {Pospieszalski}, {Poutanen}, {Prina}, {Rebolo}, {Roddis},
  {Rubi{\~n}o-Martin}, {Salmon}, {Sandri}, {Seiffert}, {Silvestri},
  {Simonetto}, {Sjoman}, {Smoot}, {Sozzi}, {Stringhetti}, {Taddei}, {Tauber},
  {Terenzi}, {Tomasi}, {Tuovinen}, {Valenziano}, {Varis}, {Vittorio}, {Wade},
  {Wilkinson}, {Winder}, {Zacchei}, \& {Zonca}}]{Bersanelli2010}
{Bersanelli} M. {et~al.}, 2010, \aap, 520, A4

\bibitem[{{Bonaldi} \& {Brown}(2014)}]{Bonaldi2014}
{Bonaldi} A., {Brown} M.~L., 2014, ArXiv e-prints

\bibitem[{{Bondi} {et~al}\mbox{.}(2003){Bondi}, {Ciliegi}, {Zamorani},
  {Gregorini}, {Vettolani}, {Parma}, {de Ruiter}, {Le Fevre}, {Arnaboldi},
  {Guzzo}, {Maccagni}, {Scaramella}, {Adami}, {Bardelli}, {Bolzonella},
  {Bottini}, {Cappi}, {Foucaud}, {Franzetti}, {Garilli}, {Gwyn}, {Ilbert},
  {Iovino}, {Le Brun}, {Marano}, {Marinoni}, {McCracken}, {Meneux}, {Pollo},
  {Pozzetti}, {Radovich}, {Ripepi}, {Rizzo}, {Scodeggio}, {Tresse},
  {Zanichelli}, \& {Zucca}}]{Bondi2003}
{Bondi} M. {et~al.}, 2003, \aap, 403, 857

\bibitem[{{Bowman}, {Morales} \& {Hewitt}(2007){Bowman}, {Morales}, \&
  {Hewitt}}]{Bowman2007}
{Bowman} J.~D., {Morales} M.~F., {Hewitt} J.~N., 2007, \apj, 661, 1

\bibitem[{{Brandt} {et~al}\mbox{.}(1994){Brandt}, {Lawrence}, {Readhead},
  {Pakianathan}, \& {Fiola}}]{Brandt1994}
{Brandt} W.~N., {Lawrence} C.~R., {Readhead} A.~C.~S., {Pakianathan} J.~N.,
  {Fiola} T.~M., 1994, \apj, 424, 1

\bibitem[{{Bull} {et~al}\mbox{.}(2015){Bull}, {Ferreira}, {Patel}, \&
  {Santos}}]{Bull2015}
{Bull} P., {Ferreira} P.~G., {Patel} P., {Santos} M.~G., 2015, \apj, 803, 21

\bibitem[{{Butler}(2002)}]{Butler2002}
{Butler} B., 2002, VLA Test Memo, 232

\bibitem[{{Cantalupo} {et~al}\mbox{.}(2010){Cantalupo}, {Borrill}, {Jaffe},
  {Kisner}, \& {Stompor}}]{Cantalupo2010}
{Cantalupo} C.~M., {Borrill} J.~D., {Jaffe} A.~H., {Kisner} T.~S., {Stompor}
  R., 2010, \apjs, 187, 212

\bibitem[{{Challinor} \& {Chon}(2005)}]{Challinor2005}
{Challinor} A., {Chon} G., 2005, \mnras, 360, 509

\bibitem[{Chang {et~al}\mbox{.}(2008)Chang, Pen, Peterson, \&
  McDonald}]{Chang2008}
Chang T.-C., Pen U.-L., Peterson J.~B., McDonald P., 2008, \pasp, 100, 091303

\bibitem[{{Chanial} \& {Barbey}(2012)}]{Chanial2012}
{Chanial} P., {Barbey} N., 2012, in SF2A-2012: Proceedings of the Annual
  meeting of the French Society of Astronomy and Astrophysics, {Boissier} S.,
  {de Laverny} P., {Nardetto} N., {Samadi} R., {Valls-Gabaud} D., {Wozniak} H.,
  eds., pp. 513--517

\bibitem[{{Chapman} {et~al}\mbox{.}(2012){Chapman}, {Abdalla}, {Harker},
  {Jeli{\'c}}, {Labropoulos}, {Zaroubi}, {Brentjens}, {de Bruyn}, \&
  {Koopmans}}]{Chapman2012}
{Chapman} E. {et~al.}, 2012, \mnras, 423, 2518

\bibitem[{{Chon} {et~al}\mbox{.}(2004){Chon}, {Challinor}, {Prunet}, {Hivon},
  \& {Szapudi}}]{Chon2004}
{Chon} G., {Challinor} A., {Prunet} S., {Hivon} E., {Szapudi} I., 2004, \mnras,
  350, 914

\bibitem[{{Church}(1995)}]{Church1995}
{Church} S.~E., 1995, \mnras, 272, 551

\bibitem[{Ciliegi {et~al}\mbox{.}(1999)Ciliegi, McMahon, Miley, Gruppioni,
  Rowan-Robinson, Cesarsky, Danese, Franceschini, Genzel, Lawrence, Lemke,
  Oliver, Puget, \& Rocca-Volmerange}]{Ciliegi1999}
Ciliegi P. {et~al.}, 1999, \mnras, 302, 222

\bibitem[{{Condon}(1974)}]{Condon1974}
{Condon} J.~J., 1974, \apj, 188, 279

\bibitem[{Condon {et~al}\mbox{.}(1998)Condon, Cotton, Greisen, Yin, Perley,
  Taylor, \& Broderick}]{Condon1998}
Condon J.~J., Cotton W.~D., Greisen E.~W., Yin Q.~F., Perley R.~A., Taylor
  G.~B., Broderick J.~J., 1998, The Astronomical Journal, 115, 1693

\bibitem[{{Danese} \& {Partridge}(1989)}]{Danese1989}
{Danese} L., {Partridge} R.~B., 1989, \apj, 342, 604

\bibitem[{{Datta}, {Choudhury} \& {Bharadwaj}(2007){Datta}, {Choudhury}, \&
  {Bharadwaj}}]{Datta2007}
{Datta} K.~K., {Choudhury} T.~R., {Bharadwaj} S., 2007, \mnras, 378, 119

\bibitem[{{Davies} {et~al}\mbox{.}(2006){Davies}, {Dickinson}, {Banday},
  {Jaffe}, {G{\'o}rski}, \& {Davis}}]{Davies2006}
{Davies} R.~D., {Dickinson} C., {Banday} A.~J., {Jaffe} T.~R., {G{\'o}rski}
  K.~M., {Davis} R.~J., 2006, \mnras, 370, 1125

\bibitem[{{Davies} {et~al}\mbox{.}(1996){Davies}, {Gutierrez}, {Hopkins},
  {Melhuish}, {Watson}, {Hoyland}, {Rebolo}, {Lasenby}, \&
  {Hancock}}]{Davies1996}
{Davies} R.~D. {et~al.}, 1996, \mnras, 278, 883

\bibitem[{{de Oliveira-Costa} {et~al}\mbox{.}(2008){de Oliveira-Costa},
  {Tegmark}, {Gaensler}, {Jonas}, {Landecker}, \&
  {Reich}}]{deOliveiraCosta2008}
{de Oliveira-Costa} A., {Tegmark} M., {Gaensler} B.~M., {Jonas} J., {Landecker}
  T.~L., {Reich} P., 2008, \mnras, 388, 247

\bibitem[{{Dickinson} {et~al}\mbox{.}(2004){Dickinson}, {Battye}, {Carreira},
  {Cleary}, {Davies}, {Davis}, {Genova-Santos}, {Grainge}, {Guti{\'e}rrez},
  {Hafez}, {Hobson}, {Jones}, {Kneissl}, {Lancaster}, {Lasenby}, {Leahy},
  {Maisinger}, {{\"O}dman}, {Pooley}, {Rajguru}, {Rebolo}, {Rubi{\~n}o-Martin},
  {Saunders}, {Savage}, {Scaife}, {Scott}, {Slosar}, {Sosa Molina}, {Taylor},
  {Titterington}, {Waldram}, {Watson}, \& {Wilkinson}}]{Dickinson2004}
{Dickinson} C. {et~al.}, 2004, \mnras, 353, 732

\bibitem[{Fomalont {et~al}\mbox{.}(2006)Fomalont, Kellermann, Cowie, Capak,
  Barger, Partridge, Windhorst, \& Richards}]{Fomalont2006}
Fomalont E.~B., Kellermann K.~I., Cowie L.~L., Capak P., Barger A.~J.,
  Partridge R.~B., Windhorst R.~A., Richards E.~A., 2006, The Astrophysical
  Journal Supplement Series, 167, 103

\bibitem[{{G{\'o}rski} {et~al}\mbox{.}(2005){G{\'o}rski}, {Hivon}, {Banday},
  {Wandelt}, {Hansen}, {Reinecke}, \& {Bartelmann}}]{Gorski2005}
{G{\'o}rski} K.~M., {Hivon} E., {Banday} A.~J., {Wandelt} B.~D., {Hansen}
  F.~K., {Reinecke} M., {Bartelmann} M., 2005, \apj, 622, 759

\bibitem[{{Gruppioni} {et~al}\mbox{.}(1999){Gruppioni}, {Ciliegi},
  {Rowan-Robinson}, {Cram}, {Hopkins}, {Cesarsky}, {Danese}, {Franceschini},
  {Genzel}, {Lawrence}, {Lemke}, {McMahon}, {Miley}, {Oliver}, {Puget}, \&
  {Rocca-Volmerange}}]{Gruppioni1999}
{Gruppioni} C. {et~al.}, 1999, \mnras, 305, 297

\bibitem[{Hamilton(2003)}]{Hamilton2003}
Hamilton J.-C., 2003, Comptes Rendus Physique, 4, 871 , dossier: The Cosmic
  Microwave Background

\bibitem[{{Hopkins} {et~al}\mbox{.}(1999){Hopkins}, {Afonso}, {Cram}, \&
  {Mobasher}}]{Hopkins1999}
{Hopkins} A., {Afonso} J., {Cram} L., {Mobasher} B., 1999, \apjl, 519, L59

\bibitem[{{Ibar} {et~al}\mbox{.}(2010){Ibar}, {Ivison}, {Best}, {Coppin},
  {Pope}, {Smail}, \& {Dunlop}}]{Ibar2010}
{Ibar} E., {Ivison} R.~J., {Best} P.~N., {Coppin} K., {Pope} A., {Smail} I.,
  {Dunlop} J.~S., 2010, \mnras, 401, L53

\bibitem[{{Jarosik} {et~al}\mbox{.}(2003){Jarosik}, {Barnes}, {Bennett},
  {Halpern}, {Hinshaw}, {Kogut}, {Limon}, {Meyer}, {Page}, {Spergel}, {Tucker},
  {Weiland}, {Wollack}, \& {Wright}}]{Jarosik2003}
{Jarosik} N. {et~al.}, 2003, \apjs, 148, 29

\bibitem[{{Kogut}(2012)}]{Kogut2012}
{Kogut} A., 2012, \apj, 753, 110

\bibitem[{{Lay} \& {Halverson}(2000)}]{Lay2000}
{Lay} O.~P., {Halverson} N.~W., 2000, \apj, 543, 787

\bibitem[{{Leach} {et~al}\mbox{.}(2008){Leach}, {Cardoso}, {Baccigalupi},
  {Barreiro}, {Betoule}, {Bobin}, {Bonaldi}, {Delabrouille}, {de Zotti},
  {Dickinson}, {Eriksen}, {Gonz{\'a}lez-Nuevo}, {Hansen}, {Herranz}, {Le
  Jeune}, {L{\'o}pez-Caniego}, {Mart{\'{\i}}nez-Gonz{\'a}lez}, {Massardi},
  {Melin}, {Miville-Desch{\^e}nes}, {Patanchon}, {Prunet}, {Ricciardi},
  {Salerno}, {Sanz}, {Starck}, {Stivoli}, {Stolyarov}, {Stompor}, \&
  {Vielva}}]{Leach2008}
{Leach} S.~M. {et~al.}, 2008, \aap, 491, 597

\bibitem[{{Liu}, {Tegmark} \& {Zaldarriaga}(2009){Liu}, {Tegmark}, \&
  {Zaldarriaga}}]{Liu2009}
{Liu} A., {Tegmark} M., {Zaldarriaga} M., 2009, \mnras, 394, 1575

\bibitem[{{Loeb} \& {Wyithe}(2008)}]{Loeb2008}
{Loeb} A., {Wyithe} J.~S.~B., 2008, Physical Review Letters, 100, 161301

\bibitem[{{Mao} {et~al}\mbox{.}(2008){Mao}, {Tegmark}, {McQuinn},
  {Zaldarriaga}, \& {Zahn}}]{Mao2008}
{Mao} Y., {Tegmark} M., {McQuinn} M., {Zaldarriaga} M., {Zahn} O., 2008, \prd,
  78, 023529

\bibitem[{{Masui} {et~al}\mbox{.}(2013){Masui}, {Switzer}, {Banavar},
  {Bandura}, {Blake}, {Calin}, {Chang}, {Chen}, {Li}, {Liao}, {Natarajan},
  {Pen}, {Peterson}, {Shaw}, \& {Voytek}}]{Masui2013}
{Masui} K.~W. {et~al.}, 2013, \apjl, 763, L20

\bibitem[{{McQuinn} {et~al}\mbox{.}(2006){McQuinn}, {Zahn}, {Zaldarriaga},
  {Hernquist}, \& {Furlanetto}}]{McQuinn2006}
{McQuinn} M., {Zahn} O., {Zaldarriaga} M., {Hernquist} L., {Furlanetto} S.~R.,
  2006, \apj, 653, 815

\bibitem[{{Mitchell} \& {Condon}(1985)}]{Mitchell1985}
{Mitchell} K.~J., {Condon} J.~J., 1985, \aj, 90, 1957

\bibitem[{{Miville-Desch{\^e}nes} {et~al}\mbox{.}(2007){Miville-Desch{\^e}nes},
  {Lagache}, {Boulanger}, \& {Puget}}]{MivilleDeschenes2007}
{Miville-Desch{\^e}nes} M.-A., {Lagache} G., {Boulanger} F., {Puget} J.-L.,
  2007, \aap, 469, 595

\bibitem[{{Murtagh} \& {Heck}(1987)}]{Murtagh1987}
{Murtagh} F., {Heck} A., 1987, Multivariate Data Analysis. Kluwer Academic
  Publishers, Boston, USA

\bibitem[{{Owen} \& {Morrison}(2008)}]{Owen2008}
{Owen} F.~N., {Morrison} G.~E., 2008, \aj, 136, 1889

\bibitem[{{Pardo}, {Cernicharo} \& {Serabyn}(2001){Pardo}, {Cernicharo}, \&
  {Serabyn}}]{Pardo2001}
{Pardo} J.~R., {Cernicharo} J., {Serabyn} E., 2001, IEEE Transactions on
  Antennas and Propagation, 49, 1683

\bibitem[{{Peterson} {et~al}\mbox{.}(2009){Peterson}, {Aleksan}, {Ansari},
  {Bandura}, {Bond}, {Bunton}, {Carlson}, {Chang}, {DeJongh}, {Dobbs},
  {Dodelson}, {Darhmaoui}, {Gnedin}, {Halpern}, {Hogan}, {Le Goff}, {Liu},
  {Legrouri}, {Loeb}, {Loudiyi}, {Magneville}, {Marriner}, {McGinnis},
  {McWilliams}, {Moniez}, {Palanque-Delabruille}, {Pasquinelli}, {Pen}, {Rich},
  {Scarpine}, {Seo}, {Sigurdson}, {Seljak}, {Stebbins}, {Steffen}, {Stoughton},
  {Timbie}, {Vallinotto}, \& {Teche}}]{Peterson2009}
{Peterson} J.~B. {et~al.}, 2009, in Astronomy, Vol. 2010, astro2010: The
  Astronomy and Astrophysics Decadal Survey, p. 234

\bibitem[{{Peterson}, {Bandura} \& {Pen}(2006){Peterson}, {Bandura}, \&
  {Pen}}]{Peterson2006}
{Peterson} J.~B., {Bandura} K., {Pen}, 2006, in Proceedings of 41st Recontres
  de Moriond

\bibitem[{{Planck Collaboration} {et~al}\mbox{.}(2014){Planck Collaboration},
  {Ade}, {Aghanim}, {Armitage-Caplan}, {Arnaud}, {Ashdown}, {Atrio-Barandela},
  {Aumont}, {Baccigalupi}, {Banday}, \& et~al.}]{PlanckCollaboration2013}
{Planck Collaboration} {et~al.}, 2014, \aap, 571, A16

\bibitem[{Platania {et~al}\mbox{.}(1998)Platania, Bensadoun, Bersanelli, Amici,
  Kogut, Levin, Maino, \& Smoot}]{Platania1998}
Platania P., Bensadoun M., Bersanelli M., Amici G.~D., Kogut A., Levin S.,
  Maino D., Smoot G.~F., 1998, \apj, 505, 473

\bibitem[{{Readhead} {et~al}\mbox{.}(2004){Readhead}, {Myers}, {Pearson},
  {Sievers}, {Mason}, {Contaldi}, {Bond}, {Bustos}, {Altamirano}, {Achermann},
  {Bronfman}, {Carlstrom}, {Cartwright}, {Casassus}, {Dickinson}, {Holzapfel},
  {Kovac}, {Leitch}, {May}, {Padin}, {Pogosyan}, {Pospieszalski}, {Pryke},
  {Reeves}, {Shepherd}, \& {Torres}}]{Readhead2004}
{Readhead} A.~C.~S. {et~al.}, 2004, Science, 306, 836

\bibitem[{{Reich} \& {Reich}(1988)}]{Reich1988}
{Reich} P., {Reich} W., 1988, \aap, 196, 211

\bibitem[{{Remazeilles}, {Delabrouille} \& {Cardoso}(2011){Remazeilles},
  {Delabrouille}, \& {Cardoso}}]{Remazeilles2011}
{Remazeilles} M., {Delabrouille} J., {Cardoso} J.-F., 2011, \mnras, 410, 2481

\bibitem[{{Remazeilles} {et~al}\mbox{.}(2014){Remazeilles}, {Dickinson},
  {Banday}, {Bigot-Sazy}, \& {Ghosh}}]{Remazeilles2014}
{Remazeilles} M., {Dickinson} C., {Banday} A.~J., {Bigot-Sazy} M.-A., {Ghosh}
  T., 2014, ArXiv e-prints (1411.3628), submitted to MNRAS

\bibitem[{{Richards}(2000)}]{Richards2000}
{Richards} E.~A., 2000, \apj, 533, 611

\bibitem[{{Sayers} {et~al}\mbox{.}(2010){Sayers}, {Golwala}, {Ade}, {Aguirre},
  {Bock}, {Edgington}, {Glenn}, {Goldin}, {Haig}, {Lange}, {Laurent},
  {Mauskopf}, {Nguyen}, {Rossinot}, \& {Schlaerth}}]{Sayers2010}
{Sayers} J. {et~al.}, 2010, \apj, 708, 1674

\bibitem[{{Seymour} {et~al}\mbox{.}(2008){Seymour}, {Dwelly}, {Moss},
  {McHardy}, {Zoghbi}, {Rieke}, {Page}, {Hopkins}, \& {Loaring}}]{Seymour2008}
{Seymour} N. {et~al.}, 2008, \mnras, 386, 1695

\bibitem[{{Shaw} {et~al}\mbox{.}(2014){Shaw}, {Sigurdson}, {Pen}, {Stebbins},
  \& {Sitwell}}]{Shaw2014}
{Shaw} J.~R., {Sigurdson} K., {Pen} U.-L., {Stebbins} A., {Sitwell} M., 2014,
  \apj, 781, 57

\bibitem[{{Smith}(1982)}]{Smith1982}
{Smith} E.~K., 1982, Radio Sci., 17, 1455Ð1465

\bibitem[{{Staggs} {et~al}\mbox{.}(1996){Staggs}, {Jarosik}, {Wilkinson}, \&
  {Wollack}}]{Staggs1996}
{Staggs} S.~T., {Jarosik} N.~C., {Wilkinson} D.~T., {Wollack} E.~J., 1996,
  \apj, 458, 407

\bibitem[{{Stompor} {et~al}\mbox{.}(2002){Stompor}, {Balbi}, {Borrill},
  {Ferreira}, {Hanany}, {Jaffe}, {Lee}, {Oh}, {Rabii}, {Richards}, {Smoot},
  {Winant}, \& {Wu}}]{Stompor2002}
{Stompor} R. {et~al.}, 2002, \prd, 65, 022003

\bibitem[{{Switzer} {et~al}\mbox{.}(2013){Switzer}, {Masui}, {Bandura},
  {Calin}, {Chang}, {Chen}, {Li}, {Liao}, {Natarajan}, {Pen}, {Peterson},
  {Shaw}, \& {Voytek}}]{Switzer2013}
{Switzer} E.~R. {et~al.}, 2013, \mnras, 434, L46

\bibitem[{{Szapudi}, {Prunet} \& {Colombi}(2001){Szapudi}, {Prunet}, \&
  {Colombi}}]{Szapudi2001}
{Szapudi} I., {Prunet} S., {Colombi} S., 2001, \apj, 548, L115ÐL118

\bibitem[{{Tatarskii}(1961)}]{Tatarskii1961}
{Tatarskii} V., 1961, {Wave Propagation in a Turbulent Medium}, Dover books on
  physics and mathematical physics. Dover

\bibitem[{{Tegmark} {et~al}\mbox{.}(2000){Tegmark}, {Eisenstein}, {Hu}, \& {de
  Oliveira-Costa}}]{Tegmark2000}
{Tegmark} M., {Eisenstein} D.~J., {Hu} W., {de Oliveira-Costa} A., 2000, \apj,
  530, 133

\bibitem[{Thompson, Moran \& Swenson(2008)Thompson, Moran, \&
  Swenson}]{thompson2008interferometry}
Thompson A., Moran J., Swenson G., 2008, Interferometry and Synthesis in Radio
  Astronomy. Wiley

\bibitem[{{Weinberg} {et~al}\mbox{.}(2013){Weinberg}, {Mortonson},
  {Eisenstein}, {Hirata}, {Riess}, \& {Rozo}}]{Weinberg2013}
{Weinberg} D.~H., {Mortonson} M.~J., {Eisenstein} D.~J., {Hirata} C., {Riess}
  A.~G., {Rozo} E., 2013, \physrep, 530, 87

\bibitem[{{White} {et~al}\mbox{.}(1997){White}, {Becker}, {Helfand}, \&
  {Gregg}}]{White1997}
{White} R.~L., {Becker} R.~H., {Helfand} D.~J., {Gregg} M.~D., 1997, \apj, 475,
  479

\bibitem[{{Wolz} {et~al}\mbox{.}(2014){Wolz}, {Abdalla}, {Blake}, {Shaw},
  {Chapman}, \& {Rawlings}}]{Wolz2014}
{Wolz} L., {Abdalla} F.~B., {Blake} C., {Shaw} J.~R., {Chapman} E., {Rawlings}
  S., 2014, \mnras, 441, 3271

\end{thebibliography}
\bsp

\appendix
\section{Atmospheric noise}

We quantify the total contribution of the atmospheric noise in Section~\ref{subsec:totalatm} and we discuss the contribution of the atmospheric turbulences induced by the water vapor inhomogeneities in Section~\ref{subsec:varatm}.

\subsection{Total atmospheric absorption and emission}\label{subsec:totalatm}
For our analysis of the effects of absorption and emission below, we consider a layer of the water vapour at a temperature $T_{\textrm{atm}}$ and an optical depth $\tau$ at a given observing frequency. Around 1\,GHz, the optical depth is dominated by oxygen with a small contribution from water, usually quantified with the precipitable water vapour (PWV). PWV is defined as the height of liquid water if all the water vapor would be condensed in a column of the atmosphere. Hence, it can undergo changes on a daily basis and between different sites. The total zenith optical depth can be calculated by the following relation at one frequency $\nu$ \citep{Butler2002}
\begin{equation}
\tau=10^{-3} \times (\tau_{\textrm{ox}}+\tau_{\text{wv}} \, \text{\textnormal{PWV (mm)}}),
\end{equation}
where $\tau_{\textrm{ox}}$ is the oxygen optical depth and $\tau_{\textrm{wv}}$ the water vapor optical depth. These parameters are constant and depend on the frequency.
We use the ATM software \citep{Pardo2001} to determine their values for an altitude of the site at 200\,m, a ground level pressure of $\sim$1000\,mbar, a water vapour scale height of 2\,km and a ground level temperature $T_{\textrm{atm}}=260$\,K, and we find $\tau_{\textrm{ox}}=7.1$, $\tau_{\textrm{wv}}=0.015$. This is consistent with other predictions in the literature \citep{Smith1982}.

The total contribution of the atmospheric emission $T_{{\textrm{em}}}$ to the power received on one receiver is related to the optical depth by
\begin{equation}\label{eq:T_em}
T_{{\textrm{em}}}=T_{\textrm{atm}}(1-e^{-\tau}).
\end{equation}
The atmosphere leads to an attenuation of the signal from a source with a flux density $S_{{\textrm{em}}}$. For an average optical depth $\tau$, the flux received by the instrument is
\begin{equation}
S_{\textrm{abs}}=S_{{\textrm{em}}}e^{-\tau}.
\end{equation}
For $\tau << 1$, we obtain $S_{{\textrm{abs}}} \sim S_{{\textrm{em}}}(1-\tau)$.
Thus, the attenuation of the source by the atmosphere is at a level of one percent. For this reason, we can consider only the atmospheric emission, which is much larger than the absorption. 

The atmospheric emission contributes to the system temperature of the instrument, adding to the noise from the emissions of the instrument subsystems and from the ground pick-up and sky (astronomical signal). In the case of a small optical depth $\tau \leq 0.05$, from Eq.~\ref{eq:T_em}, we can write $T_{{\textrm{em}}}\approx T_{{\textrm{atm}}}\tau$.  
The theoretical emission is given by
\begin{equation}
T_{{\textrm{em}}}(1 \, \text{\textnormal{GHz}}) =  1.8+3.9 \times 10^{-3} \, \text{\textnormal{PWV (mm)}}.
\end{equation}
In days with favourable atmospheric conditions PWV\,$<2$\,mm, we find $T_{{\textrm{em}}} = 1.81$\,K, and in days with bad weather PWV\,$>4$\,mm, $T_{{\textrm{em}}} = 1.82$\,K. From the models of the atmospheric emission \citep{Danese1989}, \cite{Staggs1996} predicts a similar contribution with $T_{{\textrm{em}}} = 1.7$\,K. It varies so little because $\tau_{\textrm{wv}} \ll \tau_{\textrm{ox}}$ when the weather is stable.

\subsection{Atmospheric fluctuations}\label{subsec:varatm}
The fluctuating atmospheric signal in the timelines depends on the precipitable water vapour level, the turbulence height and the thickness and the wind speed of the layer. According to the Kolmogorov model of turbulence \citep{Tatarskii1961}, we can assume that the fluctuations in water vapour occur in a turbulent layer at a height $h_{av}$ with a thickness $\delta h$, with $h_{av}>>\delta h$. The resulting power spectrum from the turbulence in the timelines is assumed to be a power-law with index $b=-11/3$ on small scales and $b=-8/3$ at large scales, and is given for both cases by \citep{Lay2000}

\begin{align}
<T^2(\alpha_x, \alpha_y)>=\frac{A}{\text{\textnormal{sin}}\, \epsilon }  \left(\frac{h_{{\textrm{av}}}}{\text{\textnormal{sin}}\, \epsilon}\right)^{5/3}\left(\alpha_x^2+ \alpha_y^2\right)^{-11/3} \nonumber \\
\text{\textnormal{if}} \;\frac{h_{\textrm{av}}}{2\Delta h \; \text{\textnormal{sin}}\, \epsilon } \ll \left(\alpha_x^2+ \alpha_y^2\right)^{1/2} \ll \alpha_{{\textrm{inner}}}, 
\end{align}
\begin{align}
<T^2(\alpha_x, \alpha_y)>=\frac{A^{'}}{\text{\textnormal{sin}}\, \epsilon }  \left(\frac{h_{\textrm{av}}}{\text{\textnormal{sin}}\, \epsilon}\right)^{5/3}\left(\alpha_x^2+ \alpha_y^2\right)^{8/3}\nonumber \\
\text{\textnormal{if}} \; \alpha_{{\textrm{outer}}}  \ll \left(\alpha_x^2+ \alpha_y^2\right)^{1/2} \ll \frac{h_{{\textrm{av}}}}{2\Delta h \; \text{\textnormal{sin}}\, \epsilon },
\end{align}
where $ \alpha_{{\textrm{inner}}}$ and $\alpha_{{\textrm{outer}}}$ are the inner and outer scales of the turbulences, $\Delta h$ is the outer scale size, and $\epsilon$ is the elevation angle. The coefficients $A$ and $A^{'}$ are the amplitudes of the turbulence. The parameters $\alpha_{x,y}$ are the angular wavenumbers, related to the spatial wavenumbers $k_{x,y}$ by $\alpha_{x,y}=k_{x,y}h_{\textrm{av}}/\text{\textnormal{sin}}\, \epsilon $. 

In order to predict the residual level of the brightness temperature fluctuations for a single dish at sea level, we follow the Church model \citep{Church1995}. We assume that the effective beam area is described by a Gaussian beam
\begin{equation}
A(x,y,z)=\frac{2\lambda^2 z^2}{\pi w^2(z)}\text{\textnormal{exp}}\left(- \frac{\-2(x^2+y^2)}{w^2(z)}\right),
\end{equation}
where $w(z)=w_0\left(1+\frac{\pi^2w_0^2}{\lambda^2z^2}\right)$.\\
The parameter $w_0$ is the beam waist radius related to the resolution $\theta_{{\textrm{FWHM}}}$ by $w_0=\lambda \sqrt{2\text{\textnormal{log}}2}/(\pi \theta_{{\textrm{FWHM}}})$.

In the limit of a long averaging time, the Church model \citep{Church1995} gives the rms fluctuations in antenna temperature for a single-dish according to the outer scale $L_0$ by
\small
\begin{equation}\label{eq:tatm}
\Delta T^2_{{\textrm{atm}}}=0.3\sqrt{\frac{\pi}{2}}L_0^{5/3}\int^{z_{\textrm{u}}}_{z_{\textrm{l}}}C^2_{\alpha}(z)T^2_{\textrm{atm}}(z)\left( 1+\frac{w^2(z)}{2L^2} \right)^{-1}\textrm{d}z,
\end{equation}
\normalsize
where $T_{\textrm{atm}}$ is the atmosphere temperature, $L=3L_0$, $z_{\textrm{u}}$ is the height of the atmosphere and $z_{\textrm{l}}$ the breakdown of the assumption $z \gg x,y$.
The parameter $C_{\alpha}$ is a measurement of the amplitude of the turbulent fluctuations and is given at 15\,GHz by
\begin{equation}
C^2_{\alpha}=2.0 \times 10^{-14} \, \text{\textnormal{m}}^{-8/3}\,\text{\textnormal{exp}}\left( -\frac{z}{z_0} \right) .
\end{equation}
To simplify the equations, we do not take into account the effect of the wind speed as done in \citet{Lay2000}. This adds a time dependence, which allows the measured fluctuations to decrease with larger integration times. 
The brightness temperature of the fluctuations induced by the water vapour molecules can be scaled to other frequencies, based on the emissivity spectrum of water vapour. The water vapour spectrum is composed of spectral emission lines overlaid on a background emission, which increases as the square of the frequency \citep{Smith1982}, so we scale Eq.~\ref{eq:tatm} as our frequency of observation to the fourth power. We assume a model of the telescope with $\theta_{{\textrm{FWHM}}}=40$\,arcmin at 1\,GHz, a temperature of the atmosphere of $T_{{\textrm{atm}}}=270$\,K and a maximum height of 2\,km. We use the value suggested in \citet{Church1995} for $L_0=10$\,m.  With these assumptions, we find $\Delta T_{{\textrm{atm}}}\sim 0.01$\,mK. 

\label{lastpage}

\end{document}